\documentclass{emulateapj}
\usepackage{rotating}
\usepackage{lscape}

\newcommand{\gprime}{$g^{\prime}$}
\newcommand{\rprime}{$r^{\prime}$}
\newcommand{\iprime}{$i^{\prime}$}
\newcommand{\Mdot}{\ensuremath{\dot M}}
\newcommand{\eff}{\ensuremath{\epsilon}}
\newcommand{\Msun}{\ensuremath{~\mathrm{M}_\odot}}

\slugcomment{Accepted by The Astrophysical Journal}

\shorttitle{Evolution of X-ray-selected AGN}
\shortauthors{Silverman et al.}

\begin{document}


\title{The luminosity function of X-ray selected Active Galactic Nuclei: Evolution of supermassive black holes at high redshift}


\author{J. D. Silverman\altaffilmark{1}, P. J. Green\altaffilmark{2}, W. A. Barkhouse\altaffilmark{3}, D.-W. Kim\altaffilmark{2}, M. Kim\altaffilmark{4},  B. J. Wilkes\altaffilmark{2}, R. A. Cameron\altaffilmark{2}, G. Hasinger\altaffilmark{1}, B. T. Jannuzi\altaffilmark{5}, M. G. Smith\altaffilmark{6}, P. S. Smith\altaffilmark{7}, H.  Tananbaum\altaffilmark{2}}

\altaffiltext{1}{Max-Planck-Institut f\"ur extraterrestrische Physik, D-84571 Garching, Germany}

\altaffiltext{2}{Harvard-Smithsonian Center for Astrophysics, 60 Garden Street, Cambridge, MA 02138}

\altaffiltext{3}{Department of Astronomy, University of Illinois, Urbana, IL, 61801}

\altaffiltext{4}{International Center for Astrophysics, Korea Astronomy and Space Science Institute, 36-1 Hwaam, Yusong, Taejon 305-348, Korea}

\altaffiltext{5}{National Optical Astronomy Observatory, P.O. Box 26732, Tucson, AZ, 85726-6732}

\altaffiltext{6}{Cerro Tololo Inter-American Observatory, National Optical Astronomical Observatory, Casilla 603, La Serena, Chile}

\altaffiltext{7}{Steward Observatory, The University of Arizona, Tucson, AZ 85721}


\begin{abstract}

We present a measure of the hard (2--8 keV) X-ray luminosity function
(XLF) of Active Galactic Nuclei up to $z\sim5$.  At high redshifts,
the wide area coverage of the {\em Chandra} Multiwavength Project is
crucial to detect rare and luminous ($L_X>10^{44}$ erg s$^{-1}$) AGN.
The inclusion of samples from deeper published surveys, such as the
{\em Chandra} Deep Fields, allows us to span the lower $L_X$ range of
the XLF.  Our sample is selected from both the hard ($z<3$;
$f_{2-8~{\rm kev}}>6.3\times10^{-16}$ erg cm$^{-2}$ s$^{-1}$) and soft
($z>3$; $f_{0.5-2.0~{\rm kev}}>1.0\times10^{-16}$ erg cm$^{-2}$
s$^{-1}$) energy band detections.  Within our optical magnitude limits
($r^{\prime},i^{\prime}<24$), we achieve an adequate level of
completeness ($>50\%$) regarding X-ray source identification (i.e.,
redshift).  We find that the luminosity function is similar to that
found in previous X-ray surveys up to $z\sim3$ with an evolution
dependent upon both luminosity and redshift.  At $z>3$, there is a
significant decline in the numbers of AGN with an evolution rate
similar to that found by studies of optically-selected QSOs.  Based on
our XLF, we assess the resolved fraction of the Cosmic X-ray
Background, the cumulative mass density of Supermassive Black Holes
(SMBHs), and the comparison of the mean accretion rate onto SMBHs and
the star formation history of galaxies as a function of redshift.  A
coevolution scenario up to $z\sim2$ is plausible though at higher
redshifts the accretion rate onto SMBHs drops more rapidly.  Finally,
we highlight the need for better statistics of high redshift AGN at
$z\gtrsim3$, which is achievable with the upcoming {Chandra} surveys.

\end{abstract}



\keywords{galaxies: active --- quasars: general --- X-rays: galaxies --- surveys}

\section{Introduction}

Our present understanding of the evolution of accreting SMBHs over
cosmic time comes from the measure of the luminosity function (i.e.,
the number undergoing a detectably luminous phase within a specific
co-moving volume as a function of luminosity and redshift) of Active
Galactic Nuclei (AGN).  Energy production through mass accretion onto
SMBHs allows us to observationally identify these sites as the
familiar AGN with Quasi-stellar objects (QSOs) the most luminous
example.  Hence, the AGN luminosity function provides a key constraint
to discern the underlying physical properties of the population (i.e.,
black hole mass and accretion rate distributions as a function of
redshift) and thereby elucidate the mechanisms (i.e., galaxy mergers
and/or self-regulated growth) that are instrumental in their formation
and evolution.

To date, an enormous effort has been undertaken to measure the
luminosity function over the wide range in luminosity spanned by AGN
out to high redshift.  The bright end has been well established to
$z\sim5$ by optical surveys \citep[e.g.,][]{ri06,cr04,wo03} which
primarily select QSOs using a multi-color photometric criteria.  The
most dramatic feature found is the rise and fall of the co-moving
space density with peak activity at $z\sim2.5$.  With an unprecedented
sample of over 20,000 QSOs in the 2dF QSO Redshift Survey (2QZ),
\citet{cr04} convincingly show a systematic decrease in luminosity
(pure luminosity evolution; PLE) from $z=2$ to the present, in
agreement with past surveys \citep[e.g.,][]{sc83,bsp88,hfc93}, which
find very few bright QSOs in the local universe.  This fading of the
luminous QSO population is thought to be due to a decrease in the mass
accretion rate \citep[e.g.,][]{ca00} that appears to be intimately
related to the order-of-magnitude decline of the cosmic star formation
rate from $z\sim1$ to the present \citep{bo98,fr99,me04b}.  The
dropoff in the space density at $z>3$ \citep{os82,wa94,sc95,fa01,wo03}
may be indicative of either the detection of the onset of accretion
onto young SMBHs or a high-redshift population that has been missed,
possibly under a veil of obscuration \citep[e.g.,][]{al05,ma05}.
Excessive amounts of dust and gas may be ubiquitous in galaxies at
early epochs due an increase in the merger rate \citep{ka07} that
induces high star formation rates \citep[e.g.,][]{ch05}.

It has been evident for quite some time that optical surveys miss a
significant fraction of the AGN population.  They fail to find the
majority of AGN due to a steeply declining luminosity function with
the low luminosity end severely affected by host-galaxy dilution.  Though current
optical selection techniques do show considerable improvement
\citep{ri05,ji06}, they still fail to account for many low luminosity
AGN.  Of equal significance, many AGN (e.g., Seyfert 2s, narrow line
radio galaxies) are missed due to dust obscuration (causing their
optical properties to differ from the type 1 QSOs) and can only be
adequately selected in the low redshift universe \citep{hao05} based
on their highly ionized, narrow emission lines.  Fortunately, AGN can
now be efficiently accounted for by selection techniques in other wave
bands such as the X-ray as demonstrated in this work, radio
\citep[e.g.,][]{wa05} and infrared \citep[e.g.,][]{po06}.  As further
elaborated below, current models based on recent observations continue
to attribute the bulk of the Cosmic X-ray Background (CXRB), the
previously unresolved X-ray emission, to these various types of
obscured AGN.

Over more than two decades, X-ray surveys have been improving and
extending the known AGN luminosity function by including sources at
low luminosity, with or without optical emission lines, and hidden by
a dense obscuring medium.  The Extended Medium Sensitivity Survey
\citep{gi90} was one of the first surveys to measure the X-ray
luminosity function (XLF thereafter) out to the QSO peak using a
sample of just over 400 AGN detected by the $EINSTEIN$ Observatory.
Since the survey only probed the more luminous AGN ($L_X>10^{44}$ erg
s$^ {-1}$) above $z=0.3$ due to the bright flux limit
($f_{0.3-3.5~{\rm keV}}>5\times10^{-14}$ erg cm$^{-2}$ s$^{-1}$), it
was quite understandable that \citet{ma91} and \citet{de92} found the
XLF to behave similarly to the optical luminosity function (i.e., PLE)
with a decreasing space density from $z\sim2$ to the present.  The
$ROSAT$ satellite with its increase in flux sensitivity
($f_{0.5-2.0~\rm{keV}}> 10^{-15}$ erg cm$^{-2}$ s$^{-1}$) enabled AGN
to be detected at lower luminosities, and out to higher redshifts
($z\sim4.5$).  Surveys ranged from the wide area and shallow $ROSAT$
Bright Survey \citep[f$_{lim}\sim10^{-12}$ erg cm$^{-2}$ s$^{-1}$;
20000 deg$^{2}$;][]{sc00} to the deep and narrow Lockman Hole
\citep[f$_{lim}\sim10^{-15}$ erg cm$^{-2}$ s$^{-1}$; 0.3
deg$^{2}$;][]{le01}.  With a compilation of 690 AGN from these fields
and other available $ROSAT$ surveys \citep{bow96,ap98,mc98,za98,ma00}
that effectively fill in the parameter space of flux and area,
\citet{mi00} were able to resolve $\approx$ 60-90\% of the soft CXRB
into point sources and generate a soft XLF that extended beyond the
QSO peak.  They found that the XLF departed from a simple PLE model,
now well known to describe X-ray selected samples as further
elaborated below.  A $luminosity$-dependent $density$ evolution (LDDE
thereafter) model was required due to the slower evolution rate of
lower luminosity AGN compared to that of the bright QSOs.  The limited
sky coverage at the faintest X-ray fluxes achievable with $ROSAT$
prevented an accurate measure of both the faint-end slope at
$z\gtrsim1$ and the overall XLF at high redshift since only a handful
of AGN were identified above a redshift of 3.

In the current era of {\em Chandra} and {\em XMM-Newton}, X-ray
surveys are now able to detect AGN and QSOs not only enshrouded by
heavy obscuration but those at high redshift ($z>3$) with greatly
improved statistics due to the superb spatial resolution and
sensitivity between 0.5 to 10 keV of these observatories.  Previous
X-ray missions such as $EINSTEIN$ and $ROSAT$ as described above were
limited to the soft band which biases samples against absorption.  The
$ASCA$ observatory \citep[e.g.,][]{ak03} successfully found many
nearby absorbed AGN but lacked the sensitivity to detect the fainter
sources contributing most of the 2--8 keV CXRB.  With deep
observations of the {\em Chandra} Deep Field North
\citep[CDF-N;][]{br01}, Deep Field South \citep[CDF-S;][]{ro02} and
Lockman Hole \citep{ha01}, a large fraction, between $\sim70\%$
\citep{wo05} and $\sim89\%$ \citep{mo03} of the hard (2--8 keV) CXRB
has been resolved into point sources.  Many of the hard X-ray sources
found so far arise in optically unremarkable bright galaxies
\citep[e.g.,][]{ba03b,to01}, which can contain heavily obscured AGN.

A more robust luminosity-dependent evolutionary scheme has emerged
from recent measures of the XLF.  With the inclusion of absorbed AGN
from {\em Chandra} and {\em XMM-Newton} surveys, lower luminosity AGN are
clearly more prevalent at lower redshifts ($z<1$) than those of high
luminosity that peak at $z\sim2.5$.  This behavior is due to the
flattening of the low luminosity slope at higher redshifts that has
been well substantiated with hard (2--8 keV) X-ray selected surveys
\citep{cow03,ue03,fi03,ba05,si05b,laf05}. Using a highly complete soft
(0.5--2.0 keV) band selected sample of over 1000 type 1 AGN,
\citet{ha05} has shown that this LDDE model accurately fits the data
and shows a gradual shift of the peak in the co-moving space density
to lower redshifts with declining luminosity.  This behavior may be
evidence for the growth of lower mass black holes emerging in an
``anti-hierarchical'' or ``cosmic downsizing'' fashion while accreting
near their Eddington limit \citep{ma04,me04a}, or the embers of a
fully matured SMBH population with sub-Eddington accretion rates.  The
former scenario has been substantiated by \citet{ba05} based on the
optical luminosities of the galaxies hosting X-ray selected AGN, and
\citet{he04} using type 2 AGN from the SDSS with the [OIII] emission
line luminosity as a proxy for the mass accretion rate and an estimate
of the black hole mass from the $M-\sigma$ relation.

Even though there has been much progress, there are remaining
uncertainties in the current measure of the XLF.  (1) A significant
number of X-ray sources in the recent surveys with {\em Chandra} and
{\em XMM-Newton} are not identified.  \citet{main05} find that the
peak of the redshift distribution shifts to higher values
($z\sim1.2-1.5$) when incorporating photometric redshifts for objects
too faint for optical spectroscopy.  (2) \citet{ba05} demonstrate that
the XLF can be fit equally well by a PLE model at $z<1.2$.  These
models only begin to substantially differ for low luminosity AGN at
$z>1.5$ where AGN statistics are quite low with most being provided by
the deep CDF-N and CDF-S observations.  New moderate depth surveys
such as the Extended {\em Chandra} Deep Field-South
\citep[E-CDF-S;][]{le05} and the Extended Groth Strip
\citep[EGS;][]{na05} will provide additional AGN at these luminosities
and redshifts but await optical followup.  (3) How does the AGN
population behave at redshifts above the peak ($z\sim2.5$) of the
optically-selected QSO population?  We have presented evidence
\citep{si04,si05b} for a similar evolution of luminous X-ray selected
QSOs to those found in the optical surveys \citep[e.g.,][]{ri06} with
a decline in the co-moving space density at $z>3$ but these AGN are
mainly type 1.  We don't expect the inclusion of luminous absorbed
QSOs to drastically alter our measure of the XLF since they may be at
most as numerous as the type 1 QSOs \citep{gi07}.  Recent radio
\citep{wa05} and near infrared \citep{br06} surveys are further
supporting a strong decline in the co-moving density at high redshift.

In the present study, we measure the XLF of AGN in the hard X-ray
(intrinsic 2--8 keV) band with an emphasis on reducing uncertainties
at high redshift ($3<z<5$).  This paper is an
extension to the preliminary results on the co-moving space density of
AGN as reported by the ChaMP survey \citep{si04,si05b}.  As previously
described, these early epochs represent the emergence of the luminous
QSOs and the rapid growth phase of young SMBHs.  To date, the limited
numbers of X-ray selected AGN at $z>3$ have constrained current
measures \citep{laf05,ba05,ue03} to lower redshifts.  Motivated by
\citet{ba05}, we use the observed soft X-ray band for AGN selection
above $z=3$ where we measure the rest-frame energies above 2 keV.  Due
to the rarity of luminous high redshift AGN, such an endeavor requires
a compilation of surveys that covers a wide enough area at sufficient
depths.  As previously mentioned, a large dynamic range from the deep,
narrow pencil beam to the wide, shallow surveys is required to measure
a luminosity function that spans low and high luminosities at a range of redshifts.
Currently, there are a handful of deep surveys with {\em Chandra} (i.e.,
CDF-N, CDF-S) and {\em XMM-Newton} (Lockman Hole) that have published
catalogs with a fair sample of low luminosity ($42<log~L_X<44$) AGN
out to $z\sim5$.  To provide a significant sample of more luminous AGN
($log~L_X>44$), the {\em Chandra} Multiwavelength Project (ChaMP) is
carrying out a wider area survey of archived {\em Chandra} fields and the
CLASXS \citep{ya04,st04} survey is imaging a contiguous area with nine
{\em Chandra} pointings.  The statistics of high redshift AGN are sure to
improve with the anticipated results from the SWIRE/{\em Chandra} (Wilkes
et al., in preparation), XBootes \citep{mu05}, E-CDF-S \citep{le05},
EGS \citep{na05}, XMM/COSMOS \citep{ha07}, and the newly approved
{\em Chandra}/COSMOS surveys.

We organize the paper as follows: Section~\ref{compile}, we describe
the various surveys used in this analysis including X-ray sensitivity,
sky area coverage, incompleteness as a function of not only X-ray flux
but optical magnitude, and AGN selection.  Our method for measuring
the luminosity function is presented in Section~\ref{methods} and the
results, including best-fit analytic models, in Section~\ref{results}
for all AGN types.  In Section~\ref{text:cxrb}, we address the
resolved fraction of the CXRB and any underrepresented source
populations.  We directly compare our luminosity function to that of
optically-selected samples in Section~\ref{text:compare_opt}.  Based
on our luminosity function, we derive in Section~\ref{history} the
accretion rate distribution as a function of redshift and the
cumulative mass density of SMBHs. Section~\ref{coevolution}, we
compare the global mass accretion rate of SMBHs to the star
formation history of galaxies out to $z\sim5$.  We end in
Section~\ref{predict} with some predictions of the numbers of high
redshift ($z>3$) AGN expected in new surveys that effectively enable
us to extend the luminosity function to these high redshifts with
accuracy.  Throughout this work, we assume H$_{\circ}=70$ km s$^{-1}$
Mpc$^{-1}$, $\Omega_{\Lambda}=0.7$, and $\Omega_{\rm{M}}=0.3$.

\section{Compilation of AGN from various X-ray surveys}

\label{compile}

\begin{figure}
\epsscale{1.1}
\plotone{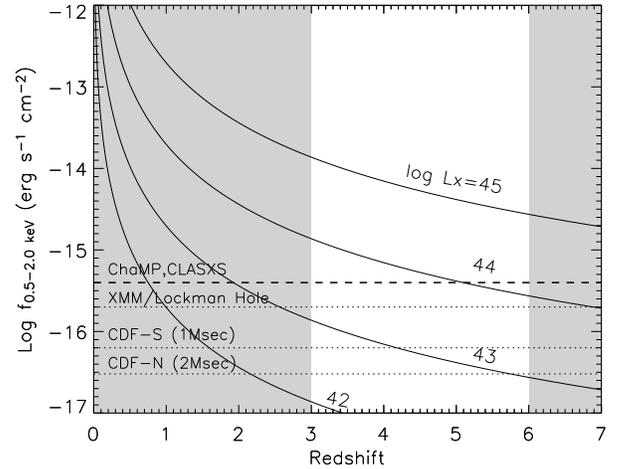}
\caption{Observed flux in soft X-rays as a function of redshift for a
theoretical AGN ($\Gamma=1.9$) of luminosity (units of erg s$^{-1}$)
as labelled.  We do not consider intrinsic absorption that may be
present in many AGN.  We highlight the redshift range $3<z<6$ over
which an extension of the known XLF is needed.  The horizontal lines
mark the flux limits of various X-ray surveys capable of detecting AGN
at these high redshifts.}

\label{model_agn}
\end{figure}

The catalog of X-ray sources from the ChaMP provides the foundation
for our measure of the AGN luminosity function.  To this, we
incorporate available catalogs from additional X-ray surveys,
described below and listed in Table~\ref{table_surveys}, that
effectively improve the coverage of the $luminosity-redshift$ plane.
In Figure~\ref{model_agn}, we illustrate the complementarity of
surveys of various depths to probe the high and low luminosity ends of
the population; the soft band is shown to highlight the feasibility of
these surveys to detect AGN out to high redshifts ($z\sim5$) by taking
advantage of the higher sensitivity of the soft band compared to
harder ($E> 2~{\rm keV}$) energy bands.  As is evident, the deep
surveys (i.e., CDF-N, CDF-S) are key to improving the quality of our
XLF at low luminosities (e.g., $log~L_X\sim42$ at $1<z<3$) thus
constraining a previously unexplored part of the luminosity function
before the {\em Chandra} and XMM-$Newton$ era.  The ChaMP and CLASXS
surveys, of shallower depth but of wider area, effectively supply the
more luminous ($log~L_X>44$) AGN especially at $z>1.5$ that are
underrepresented in the aforementioned deep fields due to a steeply
declining luminosity function.  These wider-area surveys are required
since a sky coverage of over one square degree is needed to detect
significant numbers of these rare, high redshift AGN.

We generate both a hard (2--8 keV) and soft (0.5--2.0 keV) X-ray
selected, flux-limited source catalog as motivated by \citet{ba05} to
determine their rest-frame 2--8 keV luminosity for sources with
available redshifts.  We thereby optimize uniformity in regards to
selection across the entire compilation of data sets.  The hard-band
detections are used to construct a low-redshift sample at $z<3$.  This
enables our selection to be less affected by absorbing columns up to
$N_H\sim10^{23.5}$ cm$^{-2}$.  The soft band allows us to take
advantage of {\em Chandra}'s high collecting area at low energies to
detect faint, high-redshift AGN at $z>3$.  Above this redshift,
intrinsic absorbing columns, if present, will have less effect on the
observed flux since we are detecting X-rays at rest-frame energies
above 2 keV.  This selection technique also minimizes uncertainties
induced by large $k$-corrections.

Our sample has additional selection criteria due to the dependence of
source identification on optical magnitude.  This is clearly evident
in the ChaMP by the fact that the identification of optical
counterparts is currently restricted to $i^{\prime}\lesssim23.5$
\citep{gr04,si05b}.  This is more severe ($i^{\prime}\lesssim22$) when
considering those optical counterparts that have reliable redshifts.
Additional surveys included in this study are required to have optical
imaging that is available that covers a wavelength range similar to the SDSS
\rprime ~and \iprime ~bands used in the ChaMP.  For hard-selected AGN
($z<3$), the \rprime ~band is sufficient to measure their optical
brightness.  We use the \iprime ~band for the soft-selected sample
($z>3$) since at these high redshifts extinction due to the
intervening IGM becomes significant for bandpasses at lower
wavelengths.  The addition of the {\em Chandra} Deep fields not only
pushes our overall catalog to fainter flux limits but improves our
characterization of X-ray bright, optically faint sources not yet
identified in the ChaMP.

\begin{figure}
\epsscale{1.0}
\plotone{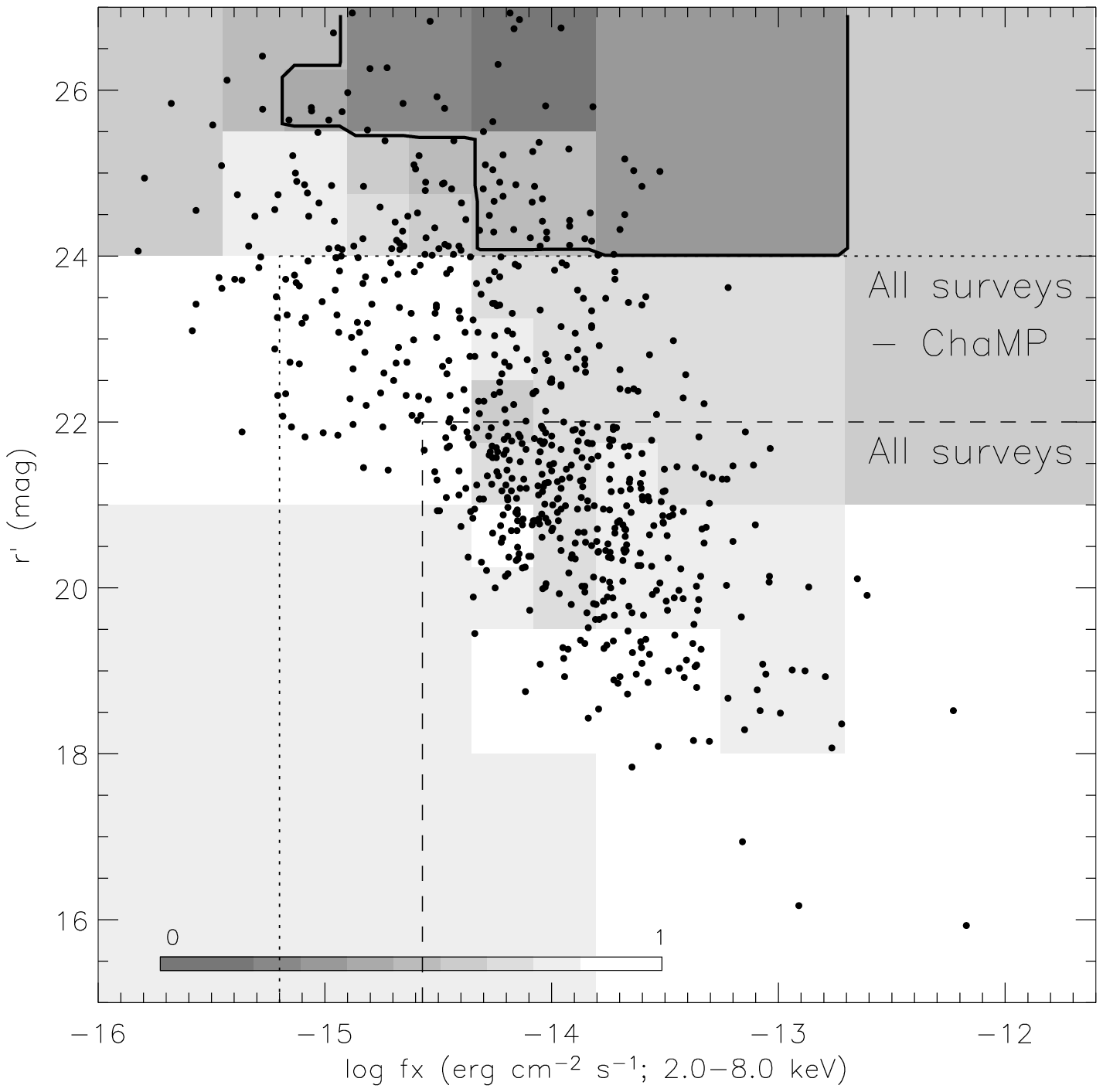}
\plotone{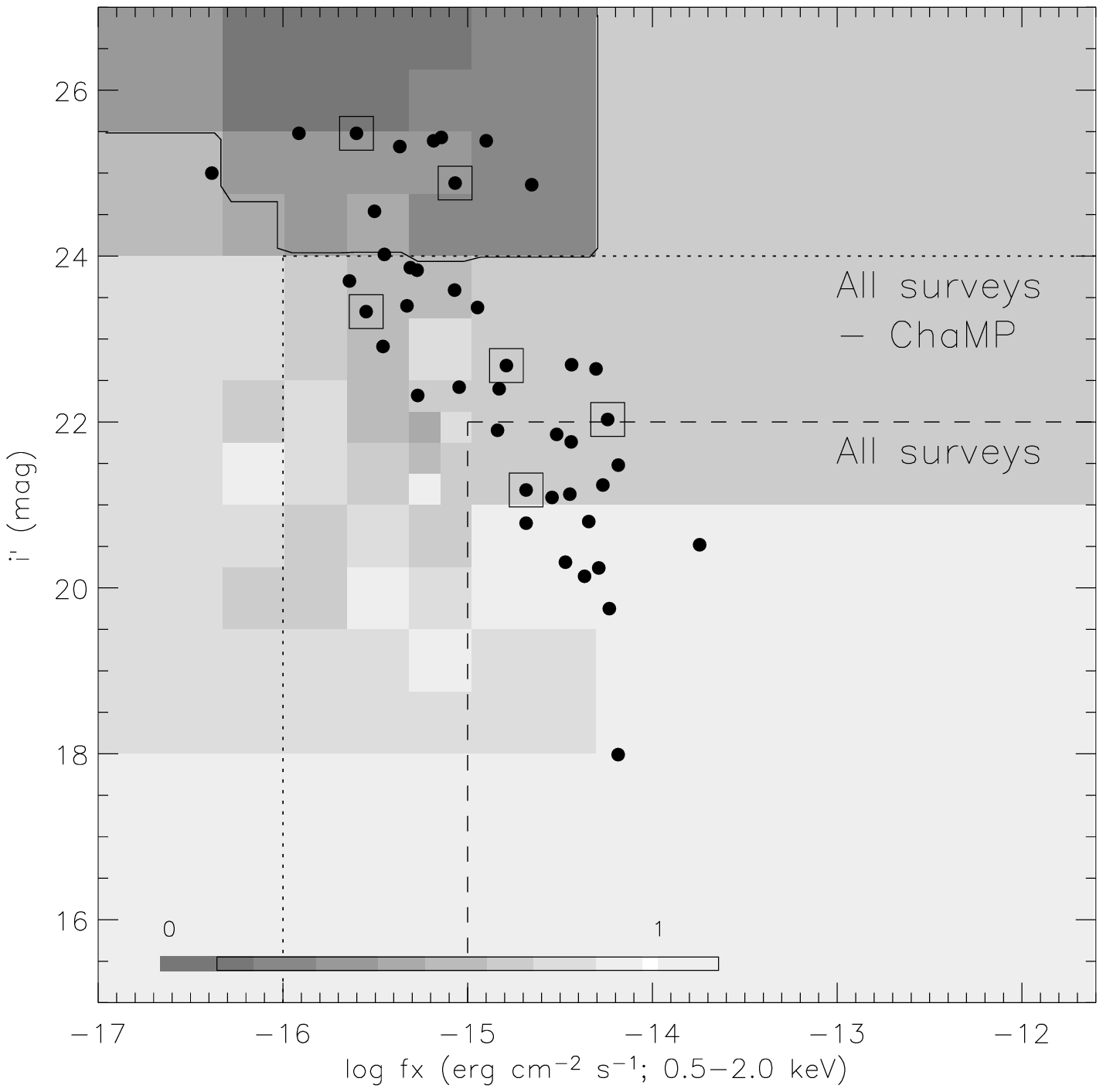}
\caption{Fraction of identified (i.e., redshift is known) sources as a
function of X-ray and optical flux in the hard/$r^{\prime}$ 
({\em top}) and soft/$i^{\prime}$ ({\em bottom})
energy bands. The grey scale image ranges from zero (darkest) to 100\%
(lightest) of sources having redshifts mainly from optical spectroscopy.
The AGN are shown with a black dot in the hard map for those at $z<3$
and soft map for those at $z>3$.  In the soft band map, those AGN at
$z>4$ are also marked by a square.  The X-ray and optical flux limits
for our analysis are shown by the dashed and dotted lines.  Since the
ChaMP survey has a reasonable degree of completeness only at brighter
flux levels, we only include ChaMP AGN that fall in the lower right
region as defined by the dashed lines.  The faintest limits (dotted
line) are determined by the fraction of identified sources ($>50\%$).}
\label{fig:selection}
\end{figure}

\begin{figure}
\epsscale{1.2}
\hspace{-1cm}
\plotone{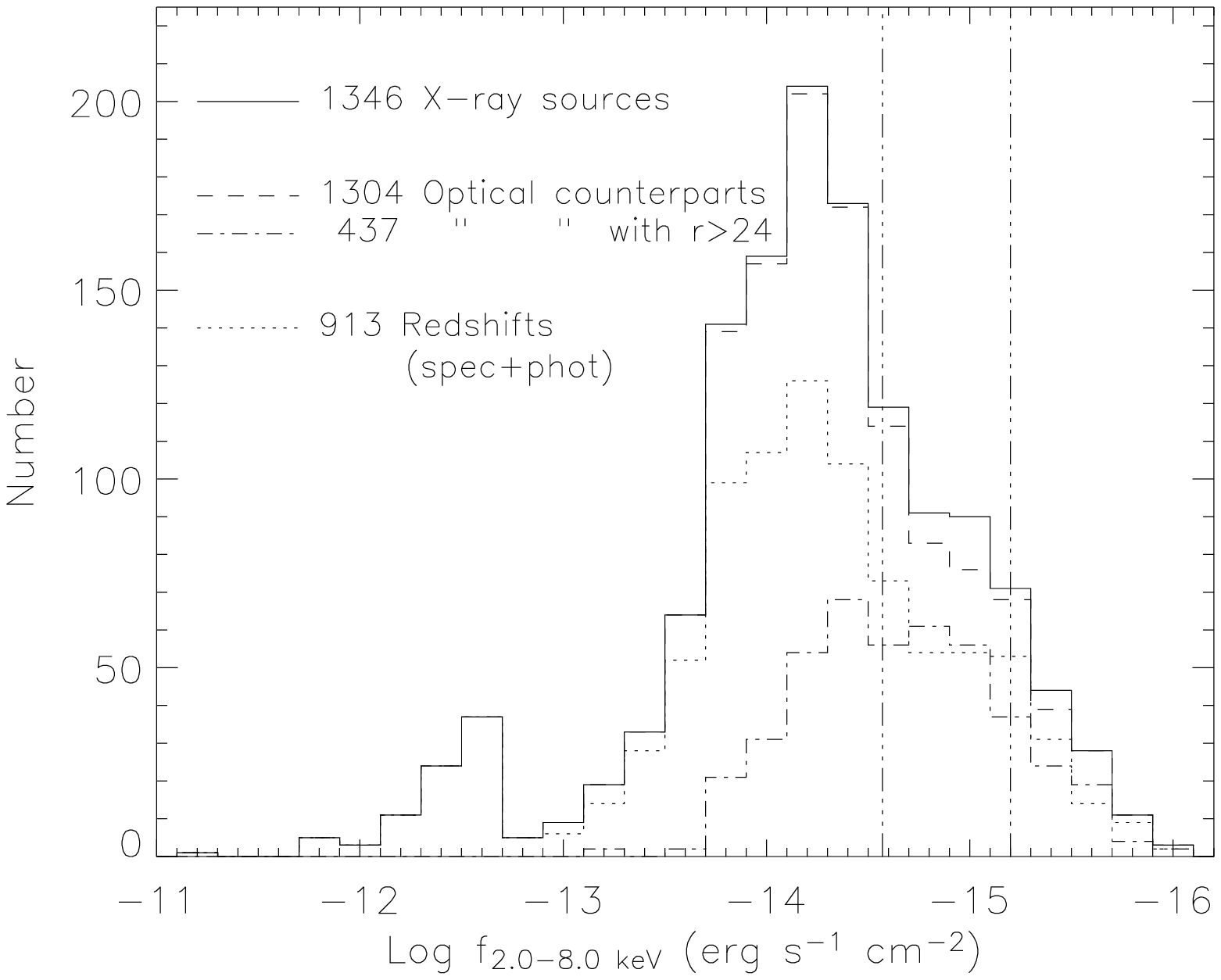}
\epsscale{1.05}
\plotone{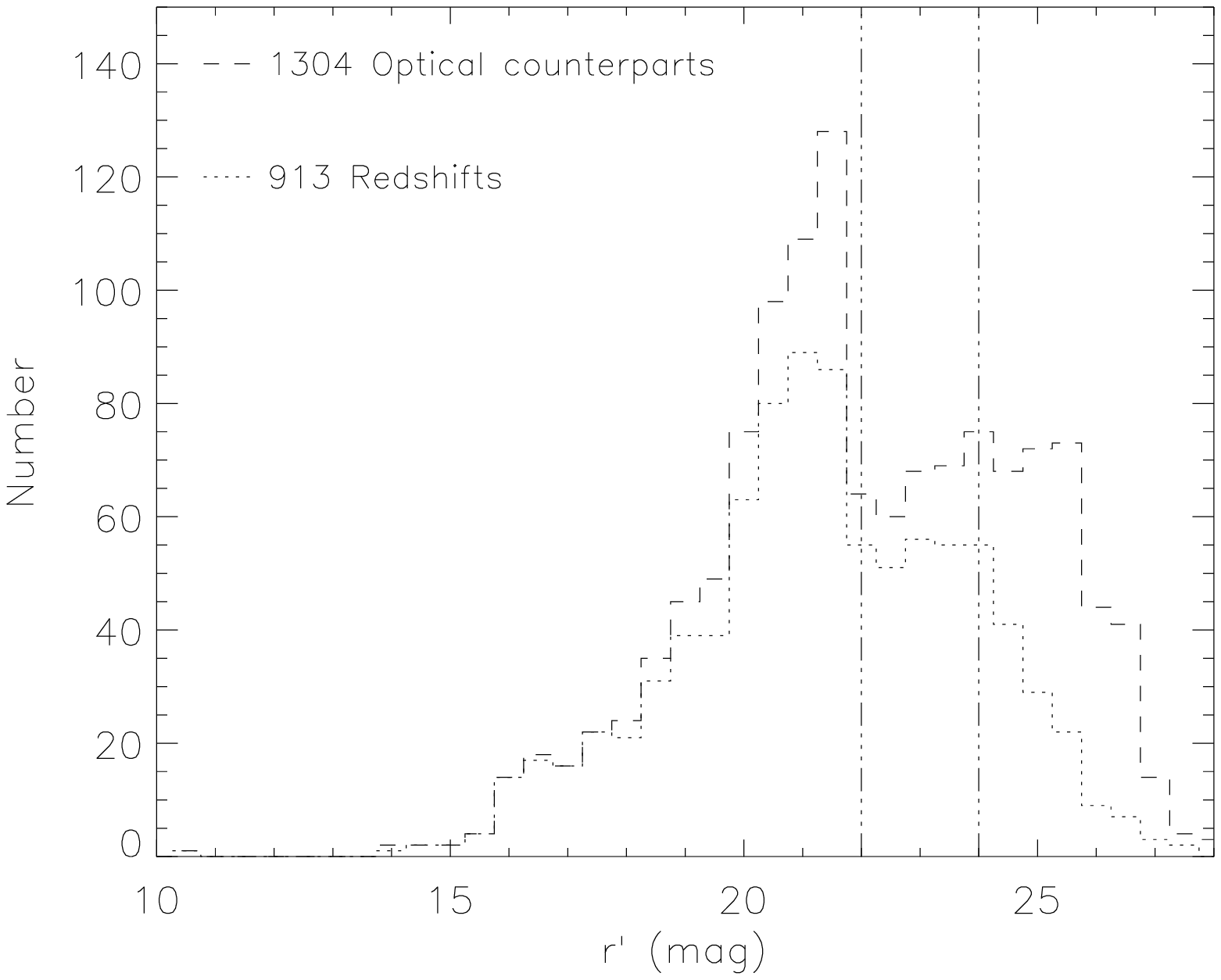}
\caption{$top$ X-ray flux distribution for the sources selected in the
2--8 keV band: all X-ray sources (solid line), those with optical
counterparts ($r^{\prime}$; dashed line), optically-faint counterparts
only ($r^{\prime}>24$; dash-dotted), and identifications
based on a spectroscopic or photometric redshift (dotted line).
$bottom$ Optical magnitude distribution.  In both plots, the vertical
lines show our chosen flux/magnitude limits with the brighter one
marking the self-imposed limit for the ChaMP survey.}
\label{fig:select_hard}
\end{figure}

\begin{figure}
\epsscale{1.1}
\plotone{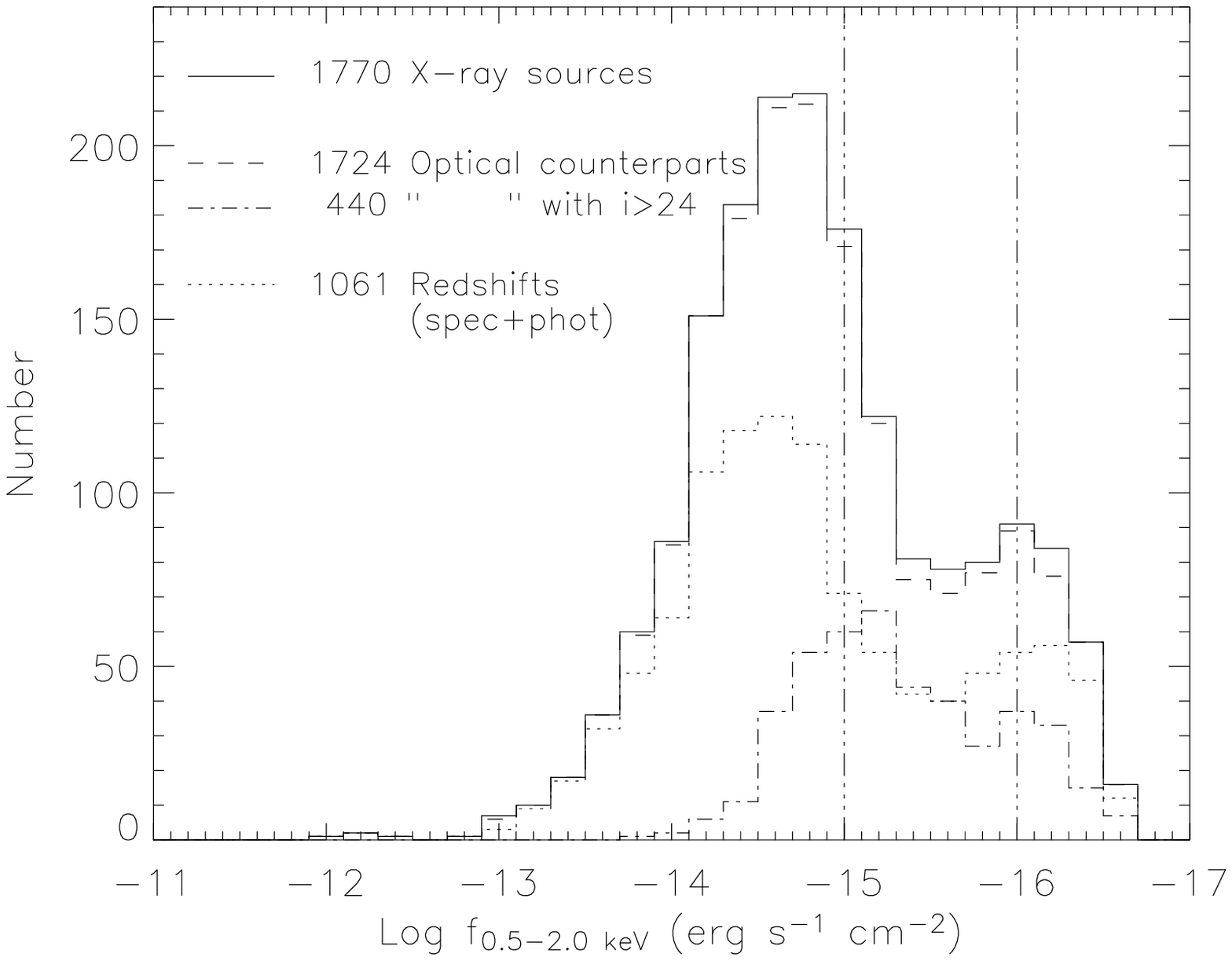}
\plotone{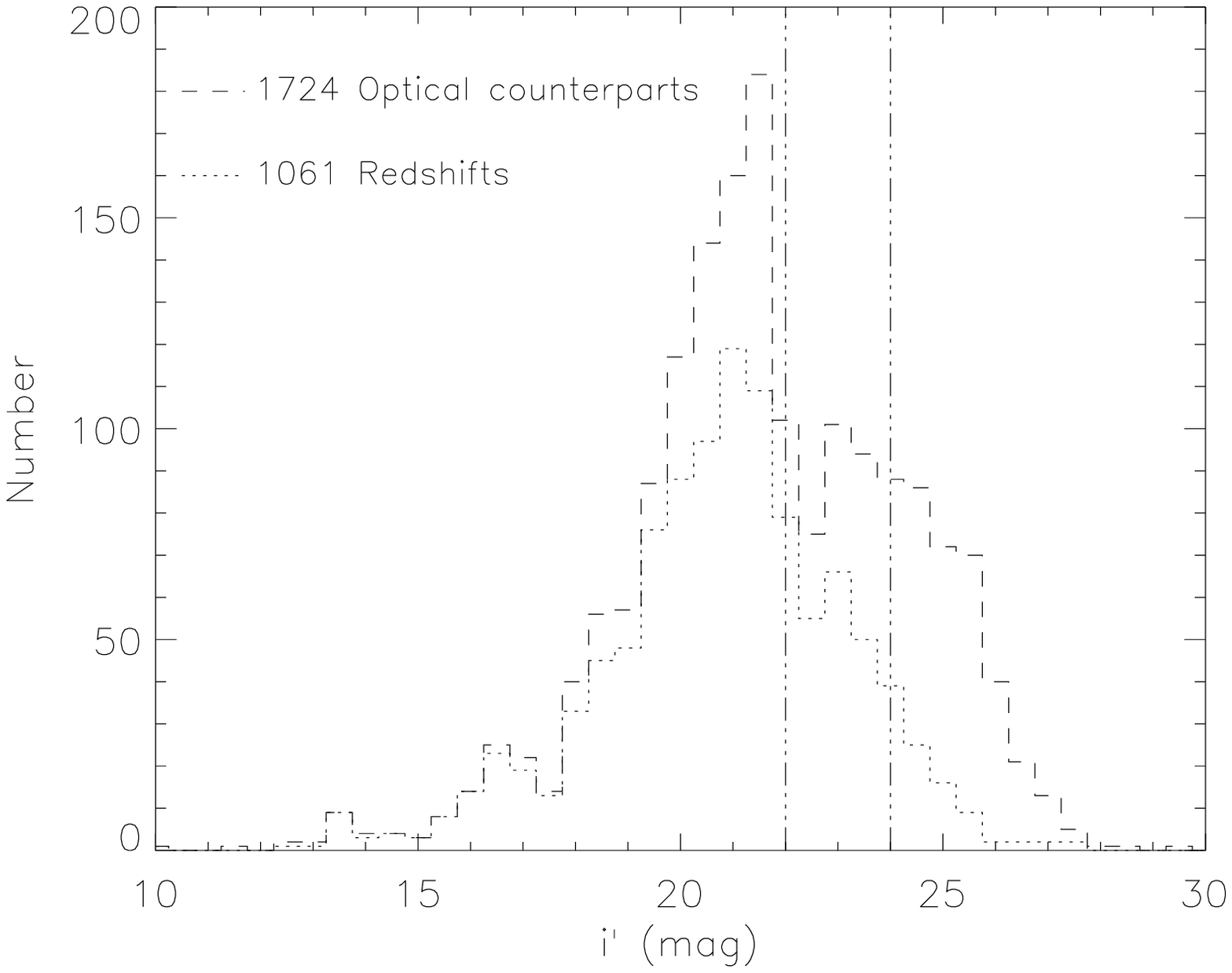}
\caption{Same as Figure~\ref{fig:select_hard} though for the soft
0.5--2.0 keV and \iprime~selected sources.}
\label{fig:select_soft}
\end{figure}

Our selection method based on both the X-ray and optical flux is
illustrated in Figure~\ref{fig:selection}.  The fraction of X-ray
sources with redshifts, either spectroscopic or photometric, is
represented as a grey scale image determined using the adaptive
binning procedure described in \citet{si05b}, and a function of both
X-ray flux and optical magnitude.  For the hard selected sample ($top$
panel), we include X-ray sources from all surveys with $f_{2-8~{\rm
keV}}>2.7\times10^{-15}$ erg s$^{-1}$ cm$^{-2}$ and $r^{\prime}<22.0$.
To this, we add sources with fainter X-ray fluxes ($f_{2-8~{\rm
keV}}<2.7\times10^{-15}$ erg s$^{-1}$ cm$^{-2}$) and optical
magnitudes ($r^{\prime}>22.0$) from all surveys with the exception of
the ChaMP and AMSS.  We set X-ray flux ($f_{2-8~{\rm
keV}}>6.3\times10^{-16}$ erg s$^{-1}$ cm$^{-2}$) and optical magnitude
limits ($r^{\prime}<24.0$) that effectively exclude the zone enclosed
by the solid contour, where $<50\%$ of the sources have redshifts.
The same sort of two zone selection ($bottom$ panel) in the $f_X$/opt
plane occurs for the soft-selected sample with the inclusion of all
sources having $f_{0.5-2~{\rm keV}}>1.0\times10^{-15}$ erg s$^{-1}$
cm$^{-2}$ and $i^{\prime}<22.0$; sources with fainter X-ray fluxes and
optical magnitudes do not include ChaMP detections since currently few
have identifications.  For the soft-selected sample, we use the
$i^{\prime}$ magnitudes to mitigate any flux loss due to extinction
blueward of Ly$\alpha$ in high-redshift AGN.  Again, we set X-ray flux
($f_{0.5-2.0~{\rm keV}}>1.0\times10^{-16}$ erg s$^{-1}$ cm$^{-2}$) and
optical magnitude limits ($i^{\prime}<24.0$) where $>$50\% of the
sources in the $f_X$/opt plane have redshifts.  The flux distribution
of X-ray sources from our above selection is shown in both the hard
(Figure~\ref{fig:select_hard}) and soft (Figure~\ref{fig:select_soft})
bands.  We also show the distribution of those with optical
counterparts and identification (i.e., redshifts) with spectroscopic
or photometric techniques.  Only the CDF-N and CDF-S surveys include
photometric redshifts due to the rich data sets in these fields.  As
is evident, we have redshifts for over $\sim50\%$ of the X-ray sources
at all fluxes.

\begin{figure*}
\epsscale{0.9}
\plotone{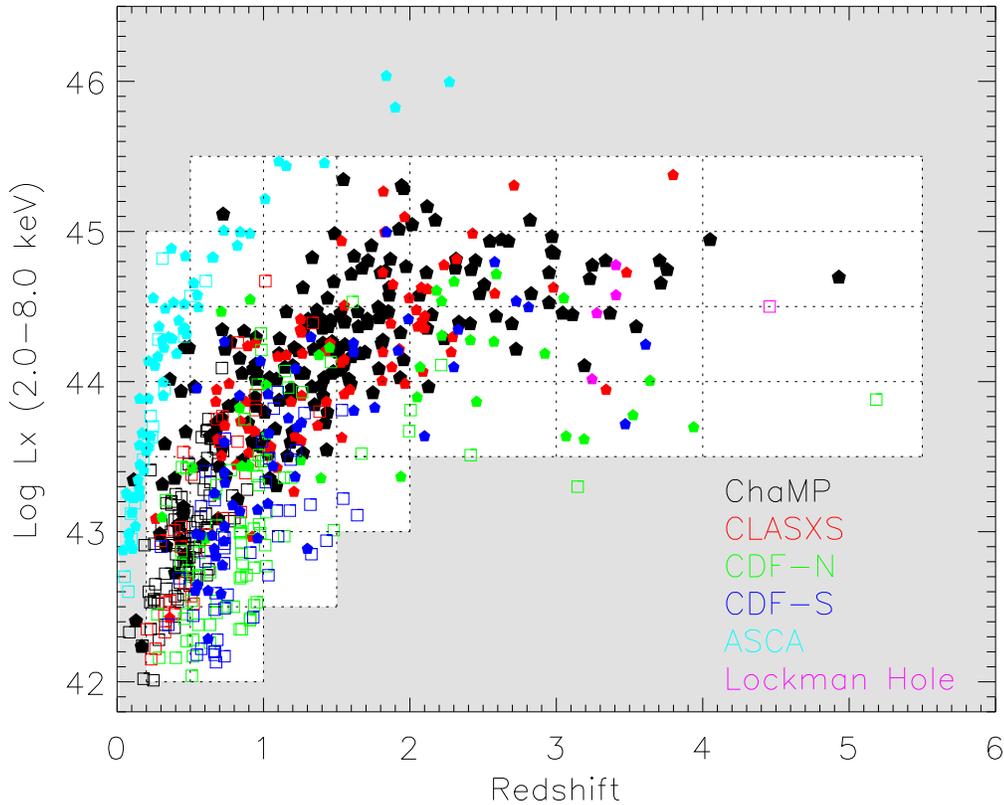}
\caption{Luminosity--redshift distribution of AGN from the X-ray surveys
as labelled.  Of the 682 AGN, 414 have broad emission lines in their
optical spectra shown by the solid pentagon.  Open squares correspond
to AGN characterized by narrow optical emission or absorption lines.  The
white area highlights the regions for which our ID fraction is
sufficient to measure the density of AGN.  The dashed lines denote
the $L_x-z$ bins used for the $1/V_a$ method.}
\label{fig:lx_z}
\end{figure*}

High X-ray luminosity becomes our single discriminant for AGN activity
since many of the traditional optical AGN signatures are not present
in obscured sources or may be outside spectroscopic coverage.  The
rest-frame 2.0--8.0 keV luminosity (uncorrected for intrinsic
absorption) is calculated by assuming a $k$-correction based on a
powerlaw X-ray spectrum with photon index $\Gamma=1.9$ and required to
exceed 10$^{42}$ erg s$^{-1}$ thereby excluding any sources with X-ray
emission primarily from a stellar or hot ISM component.  This fixed
lower limit on the luminosity ($L_{2-8~{\rm keV}}>10^{42}$ erg
s$^{-1}$) is uniformly applied to all X-ray sources from surveys
considered and yields a sample of 682 AGN with the contribution from
each survey described below and listed in Table~\ref{table_surveys}.
In Figure~\ref{fig:lx_z}, we show the luminosity--redshift
distribution of the full sample.  We now have a significant sample of
31 AGN at $z>3$ to evaluate the luminosity function at high redshift.
For the subsequent analysis, we make no attempt to subdivide the
sample into the usual optical or X-ray based categories of AGN (e.g.,
type 1/type 2, obscured/unobscured).  We do show for illustrative
purposes in Figure~\ref{fig:lx_z}, those AGN that have been classified
through spectroscopy or photometry as a type 1 AGN (filled symbol).
It is worth highlighting the fact that almost all of the AGN at $z>3$
may be optically unobscured as characterized by the presence of broad
emission lines (FWHM $>$ 1000 km s$^{-1}$) in their optical spectra.
Further spectroscopic followup with deep observations on 8m class
telescopes are required to adequately assess the contribution of
obscured AGN at these redshifts.

\subsection{{\em Chandra} Multiwavelength Project (ChaMP)}

The ChaMP\footnote{http://hea-www.harvard.edu/CHAMP/}, a survey of
serendiptious, extragalactic X-ray sources detected in fields found in
the {\em Chandra} archive, covers a large enough area at flux depths
required to detect significant numbers of AGN at high redshift.  We
refer the reader to the following ChaMP papers for full details
concerning the X-ray analysis \citep{ki07a,ki07b,ki04a,ki04b} and
optical followup \citep{gr04,si05a} programs.  Briefly, the ChaMP
point source catalog \citep{ki07a} contains 7,365 unique sources found
in 149 ACIS fields with exposure times ranging from 0.9 to 124 ksec
that corresponds to a flux limit of $f_{0.5-8.0~keV}=9\times10^{-16}$
erg cm$^{-2}$ s$^{-1}$.  The area coverage reaches $~\sim10$ deg$^2$
at the brightest fluxes.  The backbone of the optical followup program
is a deep imaging campaign with the CTIO Blanco 4m and KPNO Mayall 4m
through NOAO providing images in the SDSS filters
\gprime,~\rprime,~and \iprime~ with limiting magnitudes ($\geq24$)
depending on the depth of the X-ray exposure.  We have undertaken an
optical spectroscopic followup program utilizing the facilities
available through NOAO (i.e., WIYN/3.5m, Blanco 4m, Gemini-N) and SAO
(i.e., Magellan, MMT, FAST/1.2m).  Recently, we have begun a deeper
spectroscopic campaign on Magellan with longer exposure times ($\sim4$
hours) and Gemini-N observations to identify the optically faint
counterparts that should include the type 2 QSOs that are expected to
be found in ChaMP in greater numbers \citep[see Figure 13 of][]{si05a}
and the most distant ($z>4$) QSOs.  In total, we have accumulated
$\sim1000$ secure spectroscopic redshifts of X-ray sources to date.
In the near future, the ChaMP anticipates the release of the optical
imaging and photometry for 66 fields (W. Barkhouse et al., in
preparation) and spectroscopy products (Green et al., in preparation)
to the community.  The ChaMP is also now extending to a total of 392 
fields, using {\em Chandra} Cycle 2--6 observations overlapping the SDSS
(Green et al. 2008).  We have also begun to acquire near-infrared
observations in the J and Ks bands for select fields using ISPI on the
CTIO BLANCO 4m to measure photometric redshifts and near-infrared
properties for sources not accessible to optical spectroscopy.

\begin{figure}
\epsscale{1.0}
\plotone{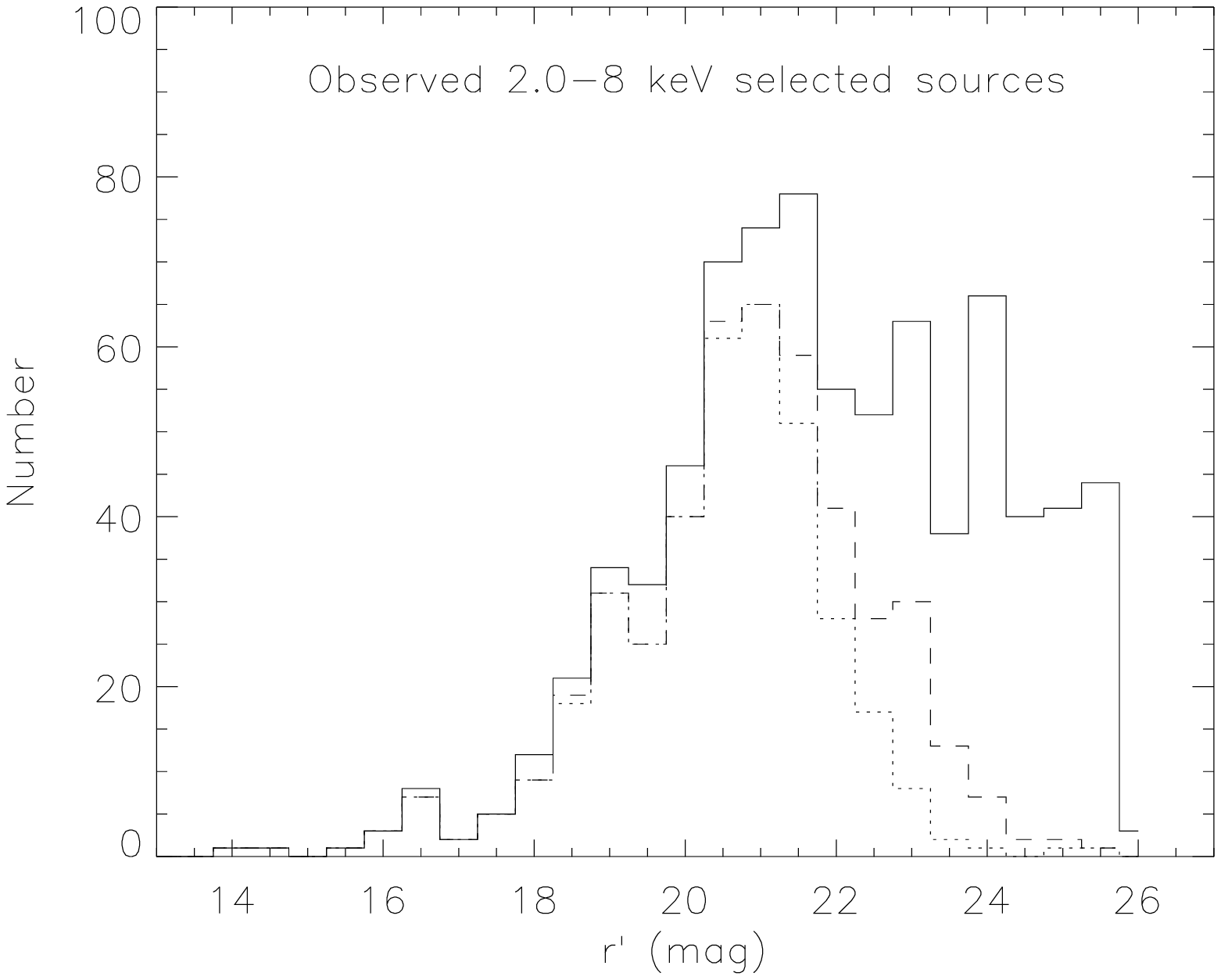}
\plotone{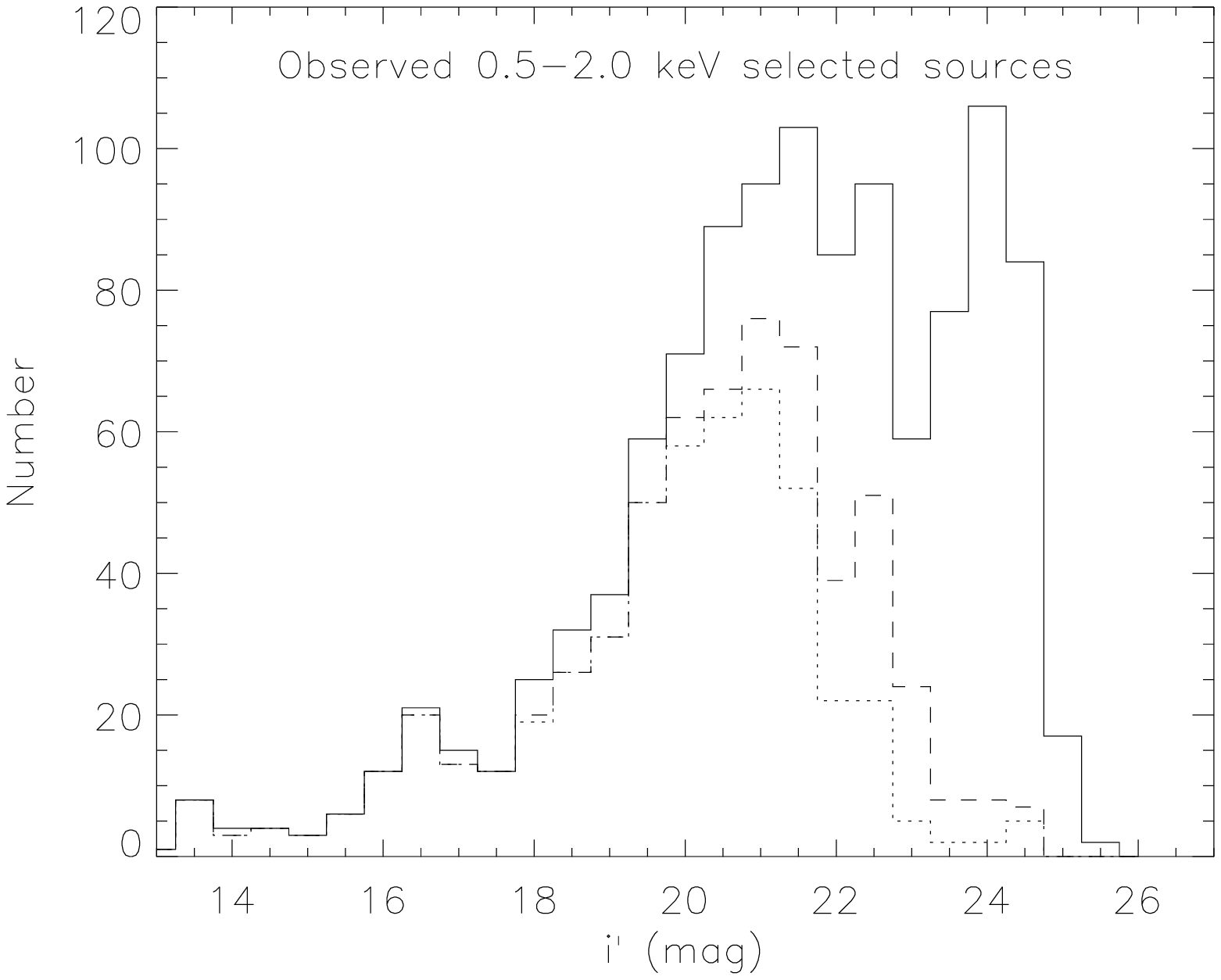}
\caption{Optical magnitude distribution (solid line) of 793 X-ray
selected sources from the ChaMP in the hard sample ($r^{\prime}$; top
panel) and 1,125 in the soft sample ($i^{\prime}$; bottom panel).  In
both panels, we show those sources that have available optical spectra
(dashed line) and those with reliable redshifts (dotted line).  See
Table~\ref{champ_stats} for a full listing of the ChaMP number
statistics.}
\label{fig:champ_stats}
\end{figure}

Based on a comprehensive catalog of 1,940 X-ray sources detected in 25
ChaMP fields (Table~\ref{champ_fields}), we have selected a subsample
of hard (2.0--8.0 keV) and soft (0.5--2.0 keV) sources subject to
X-ray and optical selection.  Fields\footnote{Two of the twenty-five
fields lack \iprime~imaging and are therefore not included in the
soft-selected sample.} were chosen to have a limiting flux capable of
detecting high redshift AGN, quality optical imaging in the
$r^{\prime}$ band for the $z<3$ sample and $i^{\prime}$ for the $z>3$
sample, and a substantial amount of optical spectroscopic followup.
Sources with greater than 10 counts in their respective selection band
are included.  We limit our sources selected in the hard band to 392
with $f_{\rm 2-8~keV}>2.7\times10^{-15}$ erg cm$^{-2}$ s$^{-1}$,
$r^{\prime}<22.0$ and the soft sample to 609 with $f_{\rm
0.5-2.0~keV}>1.0\times10^{-15}$ erg cm$^{-2}$ s$^{-1}$,
$i^{\prime}<22.0$.  These flux limits reflect the X-ray and optical
parameter space for which we have identified the majority ($\sim74\%$)
of the X-ray sources through our optical spectroscopic observations.
The brightest optical counterparts were targeted for spectroscopy.
This is clearly evident in Figure~\ref{fig:champ_stats} that shows
that most (87\%) of the optically-bright ($<22$ mag) counterparts to
both soft and hard sources have measured redshifts.  In
Table~\ref{champ_stats}, we give the number of sources for which we
have obtained optical spectra and those with reliable redshifts
(quality flag greater than 1).  A confidence level is assigned to each
redshift: 1=uncertain, 2=probably correct, 3=secure.  Spectra with an
assigned level of one tend to include BLAGN with a single low
signal-to-noise (S/N) broad emission line, whereas, those with higher
S/N can be assigned a two or three depending on the number of visible
spectral features.  Details of our spectroscopic program including
redshift measurements and quality assessment will be presented in a
future ChaMP paper (Green et al. 2008).  Spectroscopic redshifts from
the ChaMP are supplemented with an additional $\sim20$ from the SEXSI
\citep{ec06} and CYDER \citep{tr05} surveys that have carried out
observations of X-ray sources in a subset of the ChaMP fields.  We
have further removed 5 objects identified as clusters based on their
extended X-ray emission \citep[see][]{ba06}.

\begin{figure*} 
\epsscale{1.0} 
\plotone{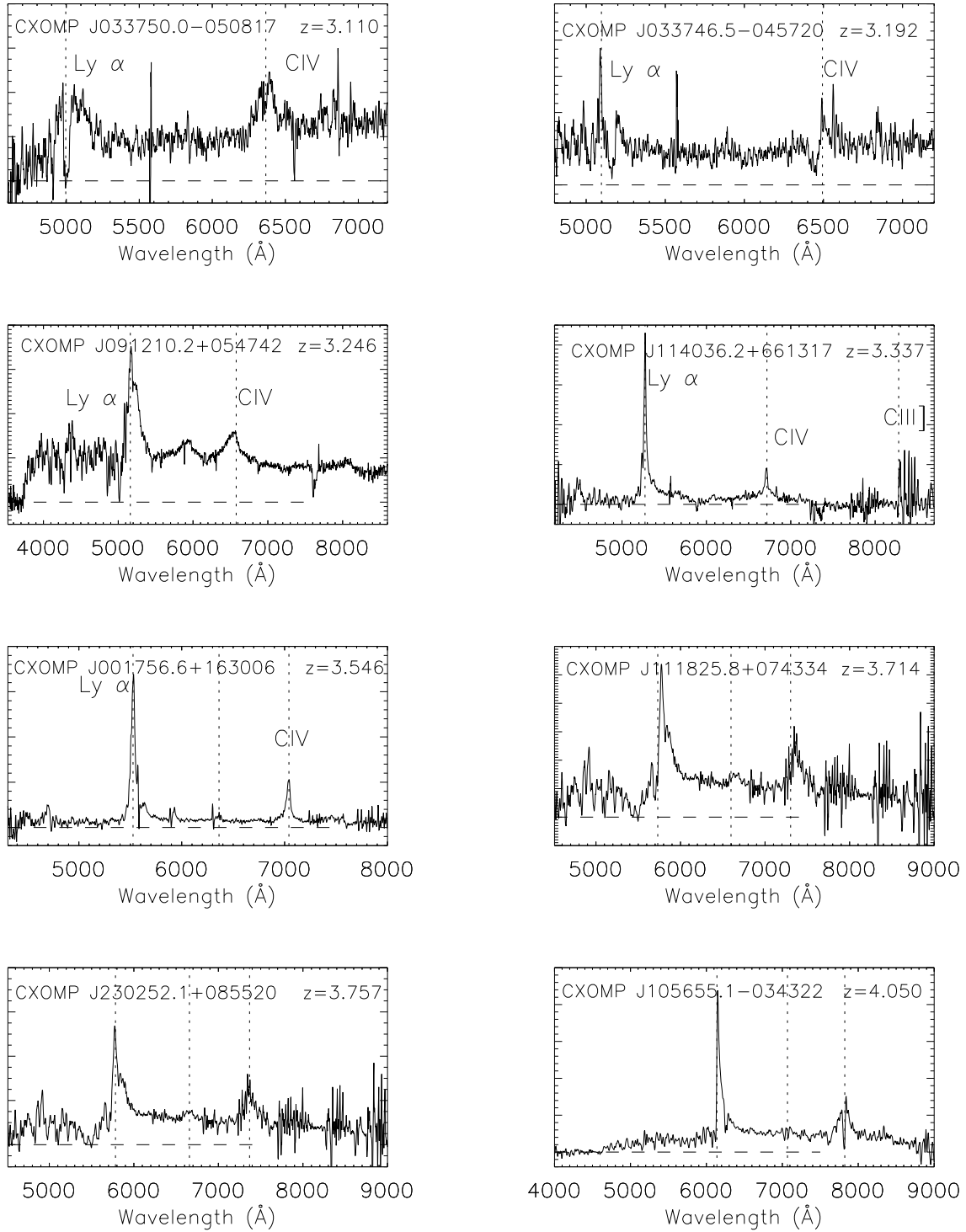} 

\caption{Optical spectra of AGN at $z>3$ identified in the ChaMP.
These comprise 8 of the 13 high redshift AGN from ChaMP included in
our analysis.  The remaining four have been published elsewhere
\citep{si02,si03}.}
\label{examples} 
\end{figure*}

The sample of 286 AGN with $L_{\rm 2-8~keV}>10^{42}$ erg s$^{-1}$ from
the ChaMP improves the statistics in select regions of the $L_X-z$
plane and increases the number of rare types as follows: (1) type 1 or
broad emission line AGN (BLAGN thereafter) at $z>1.5$ detected in the
2--8 keV band, (2) high redshift AGN, twenty-one AGN have been
identified at $z>3$ (4 at $z>4$) in the entire ChaMP program to date,
roughly half the published X-ray selected high redshift population.
Optical spectra are shown in Figure~\ref{examples} for 8 of the 13
$z>3$ AGN included in this study and not yet published), and (3)
$\sim20$ galaxies with $log~L_X\sim43$ \citep{dkim06} and no strong
optical emission lines (Equivalent width; $W_{\lambda}<5~{\rm \AA}$).

\subsection{{\em Chandra} Deep Field North (CDF-N)}

\label{cdfn}

The CDF-N, with an on-axis exposure time of $\approx2$ Ms, is the
deepest X-ray image ever obtained \citep{al03}.  Over a narrow field
(0.12 deg$^{2}$), a sample of 503 X-ray sources are detected down to a
limiting flux of $\approx2.5\times10^{-17}$ erg cm$^{-2}$ s$^{-1}$ in
the soft band (0.5--2.0 keV) and $\approx1.4\times10^{-16}$ erg
cm$^{-2}$ s$^{-1}$ in the hard band (2.0--8.0 keV).  This is
$\sim8\times$ fainter than the limiting flux of the deepest fields in
the ChaMP survey.  With follow-up observations in the radio through
ultraviolet, this combined dataset is one of the richest available for
studies of the X-ray emitting extragalactic source population.

We select sources from the X-ray catalog \citep{al03} that have
greater than 10 counts in either band to exclude those with less
certain count rates.  This restricts the X-ray catalog to 330 sources
in the hard band and 433 in the soft band.  All fluxes are corrected
for Galactic absorption using PIMMS (Version 3.6b)\footnote{PIMMS is
mostly written and maintained by Koji Mukai
(http://heasarc.nasa.gov/docs/software/tools/pimms.html)}.  The
on-axis flux limits of this sample are \hbox{$f_{\rm
2-8~keV}=1.5\times10^{-16}$ erg cm$^{-2}$ s$^{-1}$} and \hbox{$f_{\rm
0.5-2~keV}=3.06\times10^{-17}$ erg cm$^{-2}$ s$^{-1}$}.

\citet{ba03b} have identified counterparts for all of the 503 X-ray
sources with deep optical imaging
\citep[$B$,$V$,$R_C$,$I_C$,$z^{\prime}$;][]{ca04}.  We use photometric
transformations \citep{fu96,sm02} to determine \rprime~and \iprime~
magnitudes in the SDSS photometric system for optical counterparts
with $V$, $R_C$ and $I_C$ magnitudes.  Using the 486 counterparts with
magnitude in these three bands, we measure the following mean values:
$<r^{\prime}-R_C>=0.27$, $<i^{\prime}-I_C>=0.45$, and
$<R_C-I_C>=0.80$.  We use these to determine \rprime~and \iprime~when
a either $V$, $R_C$ or $I_C$ is undetermined.  Only one X-ray source
has no detections in all three bands ($V$, $R_C$ and $I_C$).  Any
uncertainty associated with these transformations does not
significantly impact our measure of the luminosity function since we
mainly consider these magnitudes to measure the fraction of identified
X-ray sources using wide magnitude bins.  Redshifts are available for
319 (74\%) of the soft-band and 223 (68\%) of the hard-band detections
with the majority ($\sim80\%$) of these obtained from optical
spectroscopy.  The remaining identifications are derived from
photometric techniques to obtain redshifts for the optically-faint
sources.  \citet{ba03b} show that these are primarily at $z>1$ and
have optical/near IR emission representative of early-type galaxies.
Above our self-imposed X-ray flux and optical magnitude limits
(Table~\ref{table_surveys}), the CDF-N sample provides 112 AGN with
\hbox{$L_{(2.0-8.0~{\rm keV})}>10^{42}$ erg s$^{-1}$} of which 8 lie
at $z>3$.

\subsection{{\em Chandra} Deep Field South (CDF-S)}

We incorporate X-ray sources detected in the 1 Msec $Chandra$
observation of the \hbox{CDF-S} \citep{gi02} with greater than 10
counts in either the soft or hard X-ray band.  Sixteen sources were
removed from the main catalog that had more than one possible optical
counterpart and four extended sources associated with an optical
group/cluster.  This results in a sample of 222 hard and 256
soft-selected X-ray sources.  The fluxes reported in \citet{gi02} are
corrected for a negligible amount of Galactic absorption.  The flux
limits of this sample are \hbox{$f_{\rm 2-8~keV}=4.6\times10^{-16}$
erg cm$^{-2}$ s$^{-1}$} and \hbox{$f_{\rm
0.5-2~keV}=5.6\times10^{-17}$ erg cm$^{-2}$ s$^{-1}$}.

We utilize the wealth of optical imaging taken of the CDF-S to
determine effective SDSS magnitudes (i.e., \rprime, \iprime) , object
types and redshifts.  The $R$ magnitude, given in \citet{gi02}, and
the relation ($<r^{\prime}-R_C>=0.27$) found using the CDF-N sources,
are used to determine \rprime.  We use the positions of the optical
counterparts of the CDF-S sources given in \citet{zh04} to search for
HST/ACS \iprime~counterparts in the GOODS data \citep{gi04}.  For
sources outside the GOODS coverage, we use the ESO imaging survey
\citep{ar01} to determine the $I$ magnitude.  In total, 301 ($88\%$)
X-ray sources listed in \citet{zh04} have an optical counterpart in
the $i_{ACS}$ or $I_{ESO}$ band.  We convert the $I_{ESO}$ to $i$
using $<i^{\prime}-I_{ESO}>=0.49$, as measured from 148 sources with
detections in both the GOODS and ESO imaging surveys.  The ESO imaging
survey provides $I$ magnitudes for 100 of the X-ray sources not
covered by the GOODS data.

Redshifts are available for 207 of the 256 soft-band sources with many
(115) from spectroscopic followup \citep{sz04}.  Multi-band photometry
confirms the redshifts of many of these sources \citep{zh04},
including a few with uncertain spectroscopic redshifts
\citep[$0<Q<2$;][]{sz04}, and identifies an additional 49
soft-selected X-ray sources.  We incorporate the classifications using
photometric techniques in cases where the quality as defined in
\citet{zh04} is $0.5\leq Q<1.0$.  This range of quality defines those
photometric redshift estimates that have similar results from at least
two of the three methods.  We also include photometric redshifts for
those sources from the COMBO-17 survey \citep{wo04} with $R<24$ since
these redshifts are fairly accurate at these brighter magnitudes.  The
CDF-S contributes 97 AGN (2 at $z>3$) with X-ray fluxes and optical
magnitudes satisfying our chosen limits (Table~\ref{table_surveys}).

\subsection{{\em Chandra} Large Area Synoptic X-ray Survey (CLASXS)}

The CLASXS survey is a contiguous mosaic of 9 {\em Chandra} ACIS-I
pointings in the low Galactic column region of the Lockman
Hole-Northwest.  The X-ray catalog \citep{ya04} contains 519 X-ray
sources detected in the 0.4--2.0 keV band with a flux above
\hbox{$5\times10^{-16}$ erg cm$^{-2}$ s$^{-1}$}.  To merge the sample
with the ChaMP, we use the Galactic corrected 0.5--2.0 keV fluxes
listed in \citet{st04} for sources with greater than or equal to 5
counts in the respective energy selection band.  We note that the
conversion from counts to flux takes into account the hardness ratio
of each individual source, similar to the \hbox{CDF-N} data.  This can
cause slight non-uniformity in our overall compilation.  We assume
that the area coverage shown in Figure 9 of \citet{ya04} in the
0.4--2.0 keV does not change significantly when using a slightly wider
energy band.  With deep optical imaging
($B$,$V$,$R_C$,$I_C$,$z^{\prime}$), \citet{st04} have identified
optical counterparts for 99\% of the X-ray sources.  As done for the
CDF-N (Section~\ref{cdfn}) sample, we use the photometric
transformations to determine \rprime~and \iprime~magnitudes in the
SDSS photometric system for optical counterparts with $V$, $R_C$ and
$I_C$ magnitudes.  Using the 451 counterparts, we measure:
$<r^{\prime}-R_C>=0.23$, $<i^{\prime}-I_C>=0.43$, and
$<R_C-I_C>=0.37$.  Only five X-ray sources have no detections in all
three $V$, $R_C$ and $I_C$ bands.  Optical spectroscopic
classifications are available \citep{st04} for 52\%, including a
significant number of optically-faint ($R>24$) counterparts.  The
spectroscopic classification scheme is nearly identical to the ChaMP.
We group the emission line objects (i.e., star forming and Seyfert 2
galaxies) and absorption line galaxies under a general ``galaxy''
category.  Above the flux limits (Table~\ref{table_surveys})
implemented for this study, there are 106 sources attributed to AGN
activity based solely on their X-ray luminosity.

\subsection{{\em XMM-Newton} Lockman Hole}

We include X-ray sources found in the deep {\em XMM-Newton}
observation of the Lockman Hole \citep{ha01} to boost our sample at
high redshift.  We only use the soft (0.5--2.0 keV) selected sources
to add to our $z>3$ sample since the optical followup has primarily
targeted soft-selected sources from the $ROSAT$ Ultra Deep Survey
\citep{le01}.  The most recent catalog (Brunner et al. in preparation)
contains 340 sources detected in the soft band with a limiting flux of
\hbox{$2\times10^{-16}$ erg cm$^{-2}$ s$^{-1}$}.  We limit the X-ray
detections to those within 12$\arcmin$ of the aimpoint since this
region contains the majority of the sources with spectroscopic
redshifts.  This sample provides a fair number of AGN at high redshift
since four have been reported \citep{le01} at $z>3$ and one additional
AGN has been identified at $z=3.244$ (Szokoly et al. in preparation).
Optical magnitudes have been converted to SDSS magnitudes as described
above (Section~\ref{cdfn}).  We further limit the catalog by selecting
only those sources with X-ray flux above \hbox{$f_{\rm
X}>1.0\times10^{-15}$ erg cm$^{-2}$ s$^{-1}$} since the fraction of
identified sources at fainter fluxes is low ($<30\%$).  Above these
flux limits, there is a reasonably high degree of completeness with
59\% (69 out of 117) of the sources identified by optical
spectroscopy.  This field adds 5 AGN at $z>3$ that also satisfies our
optical magnitude selection (\iprime $~<~$ 24).

\subsection{$ASCA$ Medium Sensitivity Survey (AMSS)}

We include AGN from the AMSS \citep{ak03} to tie down the bright end
of the luminosity function at low redshifts.  This survey presents 87
sources detected in the 2--10 keV band with fluxes down to
\hbox{$3\times10^{-13}$ erg cm$^{-2}$ s$^{-1}$}.  We convert flux to
that in the 2--8 keV band using PIMMS.  We further restrict the sample
to 76 AGN with fluxes above \hbox{$2.5\times10^{-13}$ erg s$^{-1}$
cm$^{-2}$}; this is essentially the same limit chosen by \citet{ue03}
to mitigate the influence of systematic effects near the flux limit of
the survey.  Optical photometry and source classifications are
available to easily assimilate this sample of 76 AGN into our
composite catalog.

\subsection{Sky coverage}

\begin{figure}
\epsscale{1.1}
\plotone{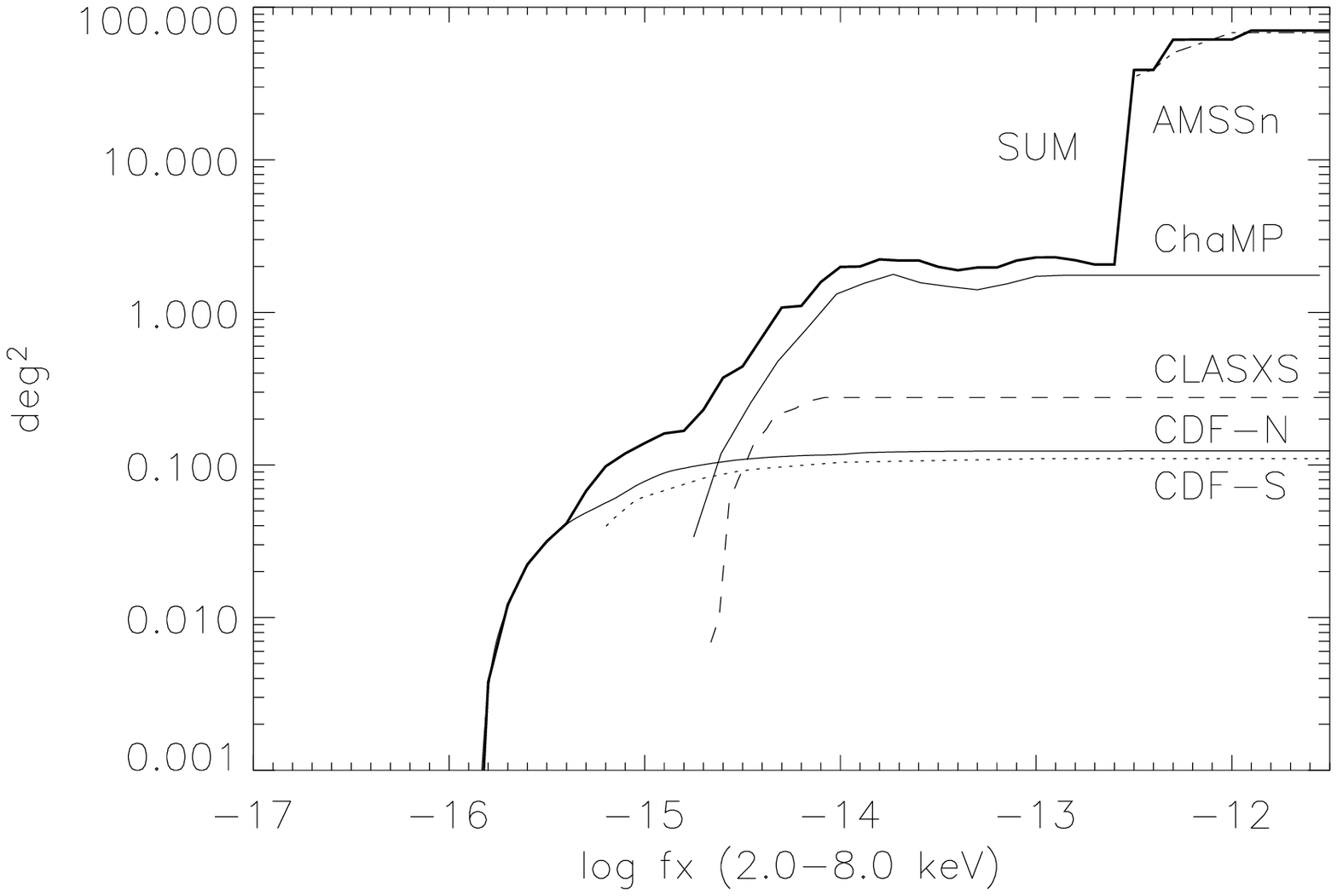}
\plotone{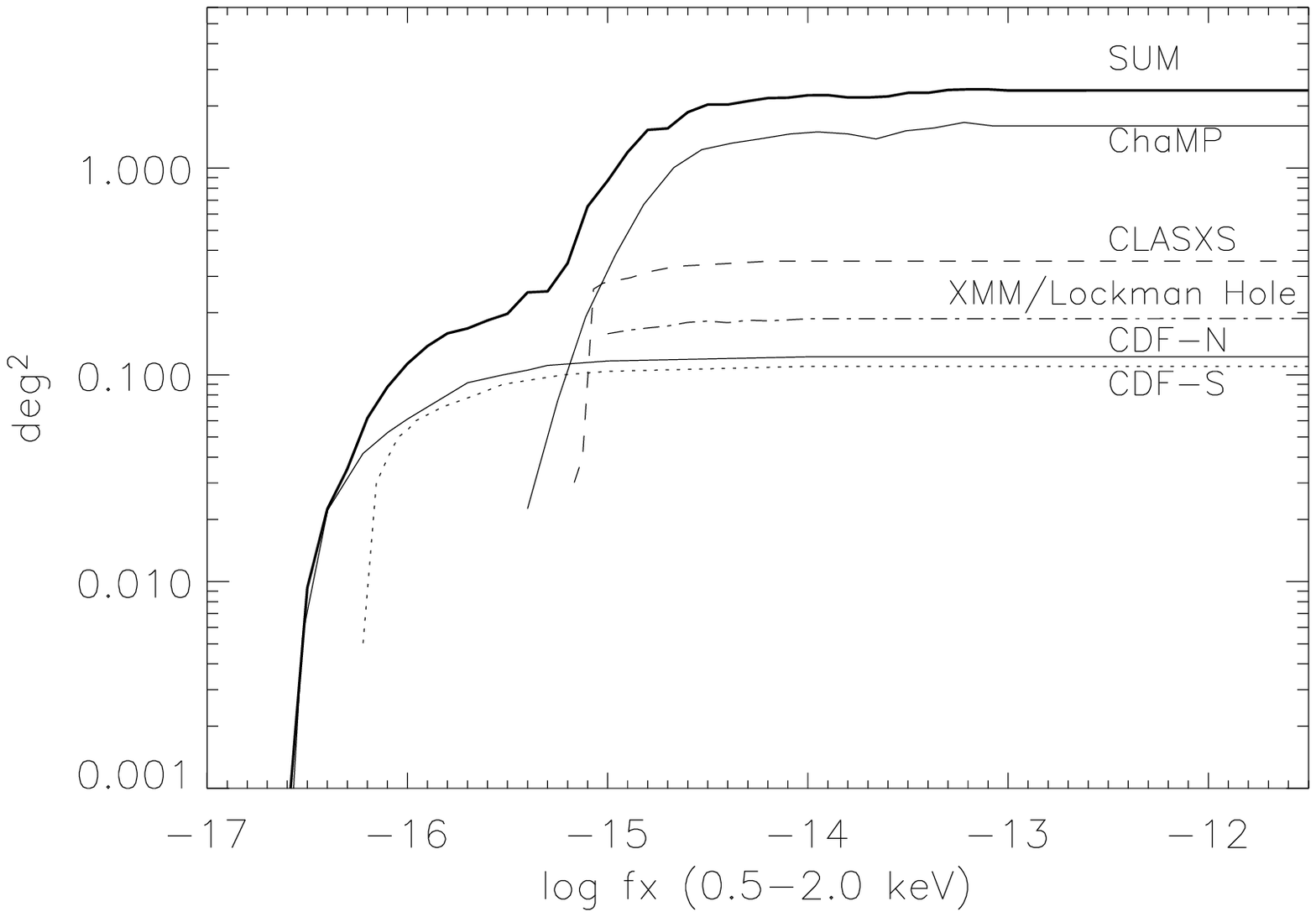}
\caption{Sky area coverage as a function of X-ray flux in the hard
(2.0--8.0 keV) band $top$ and soft band (0.5--2.0 keV) $bottom$.  The
thick line is the sum of all surveys.}
\label{fig:area}
\end{figure}

Each survey included in this study has published a measure of the sky
area as a function of flux in both the soft and hard energy bands.  To
this, we add the contribution from the ChaMP \citep{ki07a,ki07b}.  As
described briefly in \citet{si05b} and reiterated here, a series of
simulations are performed to characterize the sensitivity,
completeness, and sky area coverage as a function of X-ray flux.  The
simulations consists of three parts, 1) generating artificial X-ray
sources with MARX (MARX Technical
Manual\footnote{http://space.mit.edu/CXC/MARX}) and adding them to
real X-ray images, 2) detecting these artificial sources by using the
CIAO3.0\footnote{{\tt http://cxc.harvard.edu/ciao}} {\tt wavdetect}
and extracting source properties with XPIPE identically as performed
for actual sources, and 3) estimating the sky area coverage as a
function of flux by comparing the input and detected source
properties. The simulations are restricted to specific CCDs for ACIS-I
(I0, I1, I2, and I3) and ACIS-S (I2, I3, S2, and S3).  These CCDs are
closest to the aimpoint for each observation.  ACIS S4 is excluded
because of its high background and noise streaks.  Sources far
off-axis ($\Theta>12\arcmin$) are excluded since the flux sensitivity
is low and the PSF is degraded.  These simulations allow determination
of corrections for the source detection incompleteness at faint flux
levels quantified in the first ChaMP X-ray analysis paper
\citep{ki04b}.  In Figure~\ref{fig:area}, we show the sky coverage
from all surveys in the soft (0.5--2.0 keV) and hard (2.0-8.0 keV)
bands.  As shown, the ChaMP adds $\sim2$ square degrees at
intermediate flux levels between the narrow deep fields and the
shallower, wide area ASCA surveys.

\section{Measuring the hard (2--8 keV) XLF}
\label{methods}

We measure the differential XLF (d$\Phi/\rm dlogL$) expressed in
Equation~\ref{eq:phi_def} where $N$ is the number of AGN per unit
co-moving volume ($V$) and $log~L_{\rm x}$ as a function of X-ray
luminosity $L_{\rm x}$ and redshift $z$.

\begin{equation}
\frac{d\Phi(L_{\rm x},z)}{{\rm dlog}\,L_{\rm x}}=\frac{{\rm d}^{2}\it{N}}{{\rm d}V\,{\rm d}logL_{\rm x}}(L_{\rm x},z)
\label{eq:phi_def}
\end{equation}

\noindent This function is assumed to be continuous over the range of
luminosity and redshift spanned by our sample.  The differential
luminosity (d$logL$) is expressed as a logarithm (base 10) due to the
4 orders of magnitude covered by our sample.

Many methods to determine the luminosity function of extragalactic
populations have been implemented \citep[see][for a review]{bi88}.
The method of choice can depend on the complexity of the selection
function and/or completeness of the sample.  For example, the
traditional $1/V_{a}$ method applied to a sample of quasars
\citep{sc68} easily allows one to incorporate bivariate selection,
mainly driven by incomplete optical spectroscopic identification of
objects selected in another wavelength regime (i.e., radio, X-ray).
Alternative techniques using binned data \citep[e.g.,][]{pa99,mi01}
have been applied to X-ray selected AGN surveys, though they require
fairly complete (i.e., optical identification and redshifts) samples.
Maximum likelihood methods \citep{ma83}, though model dependent, are
most applicable when incorporating complex selection functions.  They
also mitigate effects that result from finite bin widths and irregular
sampling of each bin, typically worst near the flux limit \citep[see
Figure 1 of ][]{mi01}.

We measure the luminosity function using two methods to provide
further assurance of our results: (1) the $1/V_a$ method and (2) a
model-dependent maximum likelihood technique.  The latter allows us to
compare model fits and best-fit parameters to those published by
\citet{ue03}, \citet{ba05}, and \citet{laf05}.  In this work, we do
not consider the absorption of X-rays intrinsic to our sample thus
making no attempt to determine an N$_{\rm H}$ distribution as done in
recent studies \citep{ue03,laf05} since it is difficult to accurately
measure absorption in AGN at these high redshifts ($z>3$) due to the
very limited count statistics.  This choice does not severely affect
our results since we are using the 2--8 keV energy band and at these
high redshifts we mainly probe the bright end of the luminosity
function ($log~L_X>44$) that has been found to have a low fraction of
absorbed sources \citep[$\sim34\%$; see Figure 7 of][]{ue03}.

\subsection{Binned $1/V_a$ method}

\label{va}

We estimate the XLF in fixed luminosity and redshift bins using the
1/V$_a$ method \citep{sc68,av80} and our sample described in
Section~\ref{compile}.  For each $L-z$ bin marked in
Figure~\ref{fig:lx_z}, the value of the XLF is a sum
(Equation~\ref{eq:va}) of the contribution from each AGN falling
within this specific bin.

\begin{equation}
\frac{d\Phi(L_{\rm x},z)}{{\rm dlog}~L_{\rm x}}=\frac{1}{\Delta {\rm log} L_X}\sum_{i=1}^{N} C_i \frac{1}{V_{a,i}}
\label{eq:va}
\end{equation}

\noindent We exclude the low luminosity bins in each redshift interval
that are not well sampled at our flux limit.  Our sample at $z>3$ is
now sufficient to measure the luminosity function in two redshift
intervals.  This enables us to further probe SMBH growth at these
early epochs.

We apply a correction factor ($C_i$) for each AGN to account for the
incompleteness in our optical spectroscopic identifications.  This
factor is the reciprocal of the fraction of identified sources
($f_{\rm ID}$) at X-ray fluxes and optical magnitudes comparable to
each source, as shown by the grey scale image in
Figure~\ref{fig:selection}.  The accessible volume $V_a$ is a function
of both X-ray and optical limiting fluxes.  From \citet{ho99}, we
measure the co-moving volume for each AGN using the concordant
cosmological model for $\Omega_k=0$:

\begin{eqnarray}
V_{a,i} = \frac{c}{H_{\circ}} \int_{z_1}^{z_2}D_{L,i}^2\,\frac{(\Omega_M(1+z)^3+\Omega_{\Lambda})^{-1/2}}{(1+z)^2} \\
\nonumber \times~\Omega_{a}(L_{X,i},z)  \,dz
\label{eq:vol}
\end{eqnarray}

\noindent The solid angle $\Omega_{a}(L_{X,i},z)$ depends on the
observed flux, that varies across the width (dz) of the redshift bin,
of each AGN with specific X-ray luminosity ($L_{X,i}$) to determine an
accessible volume.  This method does not consider evolution across the
bin that may not be negligible when using large luminosity and
redshift bins.  Since we have a sample selected in two different
energy bands, the sky coverage for AGN at $z<3$ depends on their 2--8
keV X-ray flux as shown in Figure~\ref{fig:area} $top$ and their 0.5--2.0
keV flux for objects at higher redshifts ($z>3$;
Figure~\ref{fig:area} $bottom$).  Further details on our measure of $f_{\rm
ID}$ is given in \citet{si05b}.  We estimate 1$\sigma$ errors based on
a Poisson distribution due to the small number of objects per redshift
bin.

\subsection{Analytic model fitting}
\label{ml_proc}

It is highly desirable to express the luminosity function as a
smooth, well behaved analytic function in order to maximize its
versatility for a wide range of uses as later illustrated.  A well
defined model should be described by parameters that provide insight
into the physical characteristics of the population (e.g., evolution
rates).  We fit our data with three different functional forms, as
detailed below, over the full redshift ($0.2<z<5.5$) and luminosity
($42<log~L_x<46$) range spanned by our sample.  We choose models that
are conventionally used in the literature in order to directly compare
best fit parameters and to extend their applicability to higher
redshifts than have yet been accurately done in the hard X-ray band.
As is usually the case with a data set that covers different parameter
space, slight modifications of these well utilized models are
required.

To determine the values of the best-fit parameters for all models,
we implement a maximum likelihood technique \citep{ma83}. We minimize the
following expression using the MINUIT software package \citep{ja94}
available from the CERN Program Library.

\begin{eqnarray}
S=-2 \sum_{i=1}^{N}ln \frac{d\Phi(L_{\rm x}^{i},z^{i})}{{\rm dlog}~L_{\rm x}}~~~~~~~~~~~~~~~~~~~\\
\nonumber +2\int_{z_1}^{z_2}\int_{l_1}^{l_2} \frac{d\Phi(L_{\rm x},z)}{{\rm dlog}~L_{\rm x}}\Theta(f_{X},r^{\prime}/i^{\prime})  \frac{dV}{dz}dz~{\rm dlog}~L_X
\end{eqnarray}

\noindent Here, $\Theta(f_{X})$ is a correction for incomplete
redshift information as a function of X-ray flux as shown in the top
panel of Figure~\ref{fig:select_hard} and ~\ref{fig:select_soft}.  For
this exercise, we have chosen to neglect the dependence of
incompleteness as a function of optical magnitude.  After each call to
MINUIT the model is evaluated and compared to our total AGN sample
size to determine the normalization ($A_{\circ}$).  We estimate the
68\% confidence region for each parameter while allowing the other
parameters to float freely by varying the parameter of interest until
$\Delta S(=\Delta\chi^2)=1.0$ \citep{la76}.  We note that this
procedure may not accurately represent the true error interval since
we are simultaneously fitting multiple parameters.  For the
normalization ($A_{\circ}$), we use the Poisson error based on the sample
size instead of the value of $\Delta$S.

As shown by many studies \citep[e.g.,][]{ma87,bsp88}, the shape of the
XLF is best described as a double powerlaw modified by a factor(s) for
evolution.  We first implement a 'pure' luminosity evolution model
(PLE; Equations~\ref{eq:model}~--~\ref{eq:z_evol}) that is almost
identical to that used by \citet{ri06} and \citet{wo03}.  We drop the
third order term for $L_{*}$ (Equation~\ref{eq:lstar}) because any
incorporation of higher order evolution terms would require a
significantly larger sample of X-ray selected AGN not yet available.
In total, the model has 7 free parameters ($A_0$, $\gamma$1,
$\gamma$2, $L_0$, e1, e2, $z_{\rm c}$).  Best-fit results for this
model (PLE Model A) are shown in Table~\ref{ml_table} from our maximum
likelihood (ML) routine.

\begin{equation}
\frac{{\rm d} \Phi (L_{\rm X},z)}{{\rm d Log} L_{\rm X}}
= \frac{A_0}{(L_{\rm X}/L_{*}(z))^{\gamma 1} + (L_{\rm X}/L_{*}(z))^{\gamma 2}},
\label{eq:model}
\end{equation}

\noindent with

\begin{equation}
{\rm log} L_{*} = {\rm log} L_o + e1 \xi + e2 \xi^2,
\label{eq:lstar}
\end{equation}

\noindent and

\begin{equation}
\xi = {\rm log} \left ( \frac{1+z}{1+z_c} \right).
\label{eq:z_evol}
\end{equation}

We are motivated to use a model that has more flexibility to fit our
data, especially at low $L_X$.  A modified 'PLE' model
(Equations~\ref{eq:model}--\ref{eq:z_evol}) recently presented in
\citet{ho07} is quite appropriate.  This model adds two additional
free parameters, a redshift dependence to the spectral indices of the
double powerlaw.  In our case, we choose to incorporate this feature
only for the low luminosity slope (Equation~\ref{eq:gamma}) that
evidently changes with redshift.  Best-fit results for this model
(mod-PLE/Model D) are shown in Table~\ref{ml_table} from our ML
routine.

\begin{equation}
\gamma 2=(\gamma 2)_o \left(\frac{1+z}{1+zc}\right)^{\alpha}
\label{eq:gamma}
\end{equation}

Third, we implement a LDDE
(Equations~\ref{eq:ldde1}~--~\ref{eq:ldde3}; Note that for this case
$L_*=L_0$) model that has been useful in quantitatively describing the
shape and evolution of the luminosity function of X-ray selected AGN
not only in the hard band \citep{ue03,laf05} but the soft band
\citep{ha05} as well. \citet{mi00} introduced a slight variant of this
LDDE model to fit the AGN sample from a compilation of $ROSAT$ surveys
of varying depth and area coverage.  Using this model, they were able
to provide the first evidence that a luminosity-dependent evolution
scheme is required for X-ray selected samples.

\begin{equation}
\frac{{\rm d} \Phi (L_{\rm X}, z)}{{\rm d Log} L_{\rm X}}= \frac{{\rm d} \Phi (L_{\rm X}, 0)}{{\rm d Log} L_{\rm X}} e(z, L_{\rm X})
\label{eq:ldde1}
\end{equation}

\noindent where

\begin{equation}
e(z,L_{\rm x}) = \left\{
        \begin{array}{ll}
        (1+z)^{e1} & (z \leq z_{\rm c}) \\
        e(z_{\rm c})[(1+z)/(1+z_{\rm c})]^{e2} & (z > z_{\rm c})\\
        \end{array}
       \right. ,
\label{eq:ldde2}
\end{equation}

\noindent and

\begin{equation}
z_{\rm c}(L_{\rm x}) = \left\{
        \begin{array}{ll}
        z_{\rm c}^{*}(L_{\rm x}/L_{\rm a})^\alpha &
        (L_{\rm x} \leq L_{\rm a}) \\
        z_{\rm c}^{*} & (L_{\rm x}> L_{\rm a})\\
        \end{array}
       \right. .
\label{eq:ldde3}
\end{equation}

In terms of model parameters, we aim to extend the applicability of
this model out to $z\sim5$ with a re-evaluation of the value of $e2$
that describes the evolution rate beyond the cutoff redshift $z_c$.
From our preliminary measure of the co-moving space density of AGN
\citep{si05b}, we expect this value to differ sharply from other
studies \citep{ue03,laf05} with a much stronger negative evolution
rate that is similar to those found in optical QSO studies
\citep[e.g.,][]{wa94,sc95,wo03,ri06}.  In addition, the values of
$\alpha$ and $z_{\rm c}^{*}$ may differ due to our larger AGN sample
at high redshift.  To further simplify this complex model, we fix
$L_{\rm a}$ to the value used by \citet{ue03}.  In total there are 8
free parameters($A_0$,$\gamma$1, $\gamma$2, $L_0$, e1, e2, $z_{\rm
c}^{*}$, $\alpha$).  Best-fit results for this model (LDDE Model B and
C) are shown in Table~\ref{ml_table} from our ML routine.

\clearpage

\section{Hard XLF: Results}
\label{results}
\subsection{$1/V_a$ method}

\begin{sidewaysfigure*}
\centering
\includegraphics[angle=90,scale=0.9]{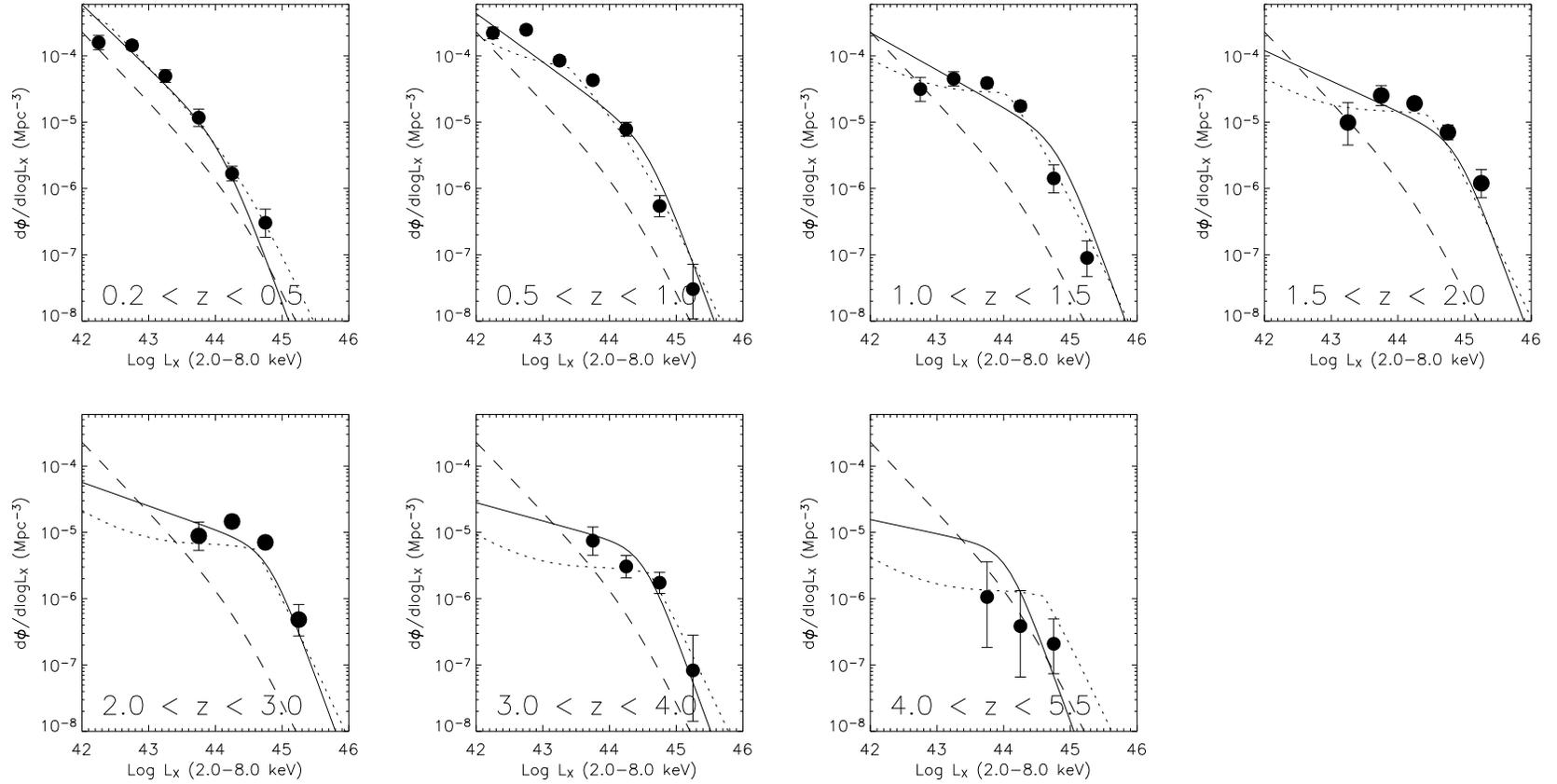}
\caption{X-ray luminosity function.  The results from the binned
$1/V_a$ method are given by filled circles with 1$\sigma$ errors.  Our
best-fit analytic models using an unbinned maximum likelihood method
are shown by the lines (solid=mod-PLE-Model D; dotted=LDDE-Model C).
The dashed line is our LDDE model C evaluated at $z=0$.}
\label{fig:xlf}
\end{sidewaysfigure*}

Our measure of the XLF, using methods detailed in the previous
section, is presented in Figure~\ref{fig:xlf} for seven redshift
intervals with bins of finite width in luminosity.  We show smooth
analytic model curves (see following section for details) including
that evaluated at $z=0$ (dashed line) to aid in our visualization of
the evolution with redshift and luminosity.  First, we see that the
shape of the XLF up to $z\sim 2$ is well approximated by the familiar
double powerlaw with a steeper bright end slope.  The shape at higher
redshifts is less constrained due to the limited statistics especially
below the break or knee (i.e., characteristic luminosity dividing the
faint and bright end slopes).  To first order, there is an overall
shift of the XLF to higher luminosities as a function of increasing
redshift (i.e., pseudo-PLE) up to $z\sim3$ as reported by many studies
in various wavebands over the past few decades.  There is a decline in
either luminosity or space density \citep{ba05,ha05,si05b} at higher
redshifts similar to the behavior seen in optical
\citep[e.g.,][]{fa01,wo03,ri06} and radio-selected \citep{wa05}
samples.  Our data preclude us from distinguishing these two modes of
evolution at $z>3$.  Second, we see that there is evidence for a
flattening of the faint-end slope with increasing redshift most
noticeable by comparing the data points with the $z=0$ model (dashed
line) in the redshift interval $1.0<z<1.5$.  The need to utilize more
complex models (i.e., LDDE), when dealing with X-ray selected samples
\citep{ha05,ue03,mi00}, is primarily due to this behavior.

\begin{figure*}
\includegraphics[angle=90,scale=0.7]{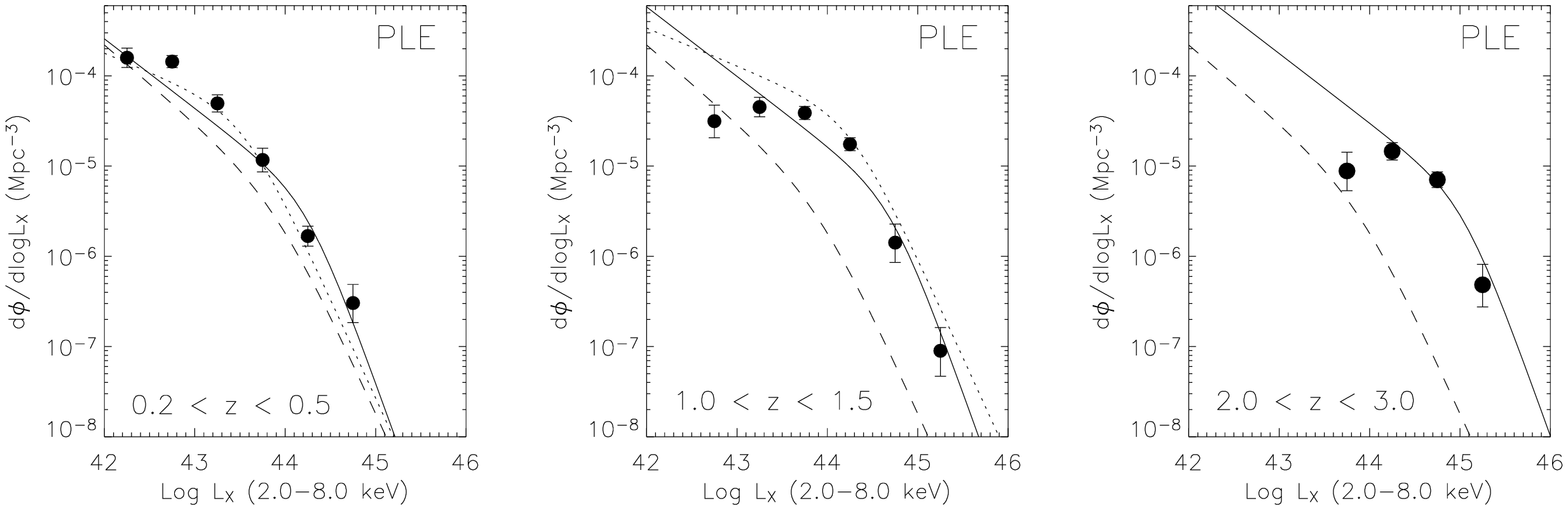}
\includegraphics[angle=90,scale=0.7]{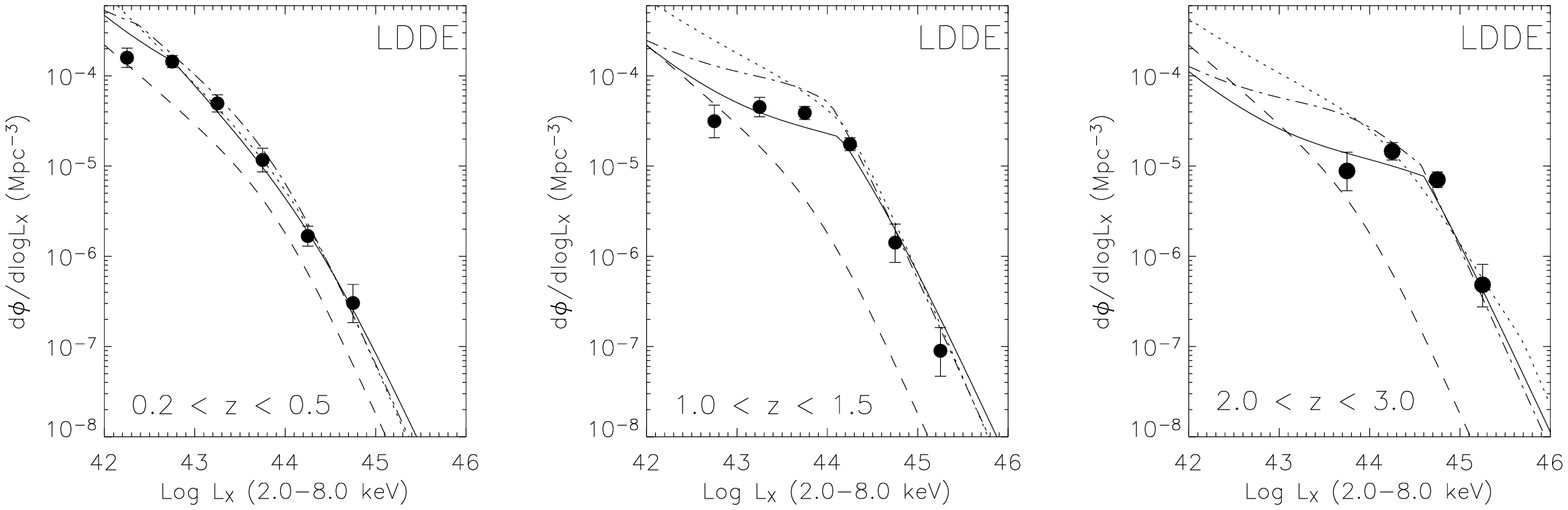}
\caption{Comparison of our binned XLF ($1/V_a$ method; filled points)
and analytic model fits at $z<3$ to recently published models. $Top$
ChaMP PLE model A (solid line) in three redshift intervals compared to
\citet{ba05} (dotted line) in the first two panels.  $Bottom$ Our best fit
LDDE model B (solid line) compared to \citet{ue03} (dashed-dotted
line) and \citet{laf05} (Fit \#4; dotted line).  The dashed line is
the Ueda model evaluated at $z=0$ in all panels.}
\label{xlf_comp}
\end{figure*}

In Figure~\ref{xlf_comp} we compare our binned measure of the XLF to
the best-fit PLE \citep{ba05} and LDDE \citep{ue03,laf05} models to
check for consistency at $z<3$.  \citet{ba05} report that a PLE model
best represents the data at $z<1.2$, with characteristic parameters
including evolution rates similar to optically-selected samples
\citep[e.g.,][]{cr04}.  As shown in the two low redshift panels
($top$; $z<1.5$), it is apparent that the PLE model provides an
adequate representation of the shape and evolution of the population.
As previously mentioned, a flattening of the faint-end slope, most
evident at $1.0<z<1.5$ (top, middle panel), is apparent and cannot be
accounted with a PLE model.  Our sample does not provide enough AGN
with luminosities below the knee and $z>2$ (top, right panel) to
constrain the faint-end slope; this an unfortunate consequence of our
optical magnitude limits and incomplete spectroscopic followup.  In
the lower panels, we compare our data with the more complex LDDE of
\citet{ue03} and \citet{laf05}.  In general, the data agree well with
both models that are nearly equivalent especially at high
luminosities.  It is worth noting that there is some slight
uncertainty in the slope and normalization at the faint end
($log~L_{X}<44$) that may be due to X-ray sources that haven't been
included due to our optical magnitude selection.  We note that the
lowest luminosity point in each of the panels is clearly affected by
the fact that AGN do not entirely fill each $L_{X}-z$ bin (see
Figure~\ref{fig:lx_z}) that falls on or near our flux limits.  An
accurate measure of the faint-end XLF is beyond the scope of this
paper and requires more complete samples.

\begin{figure}
\epsscale{1.1}
\plotone{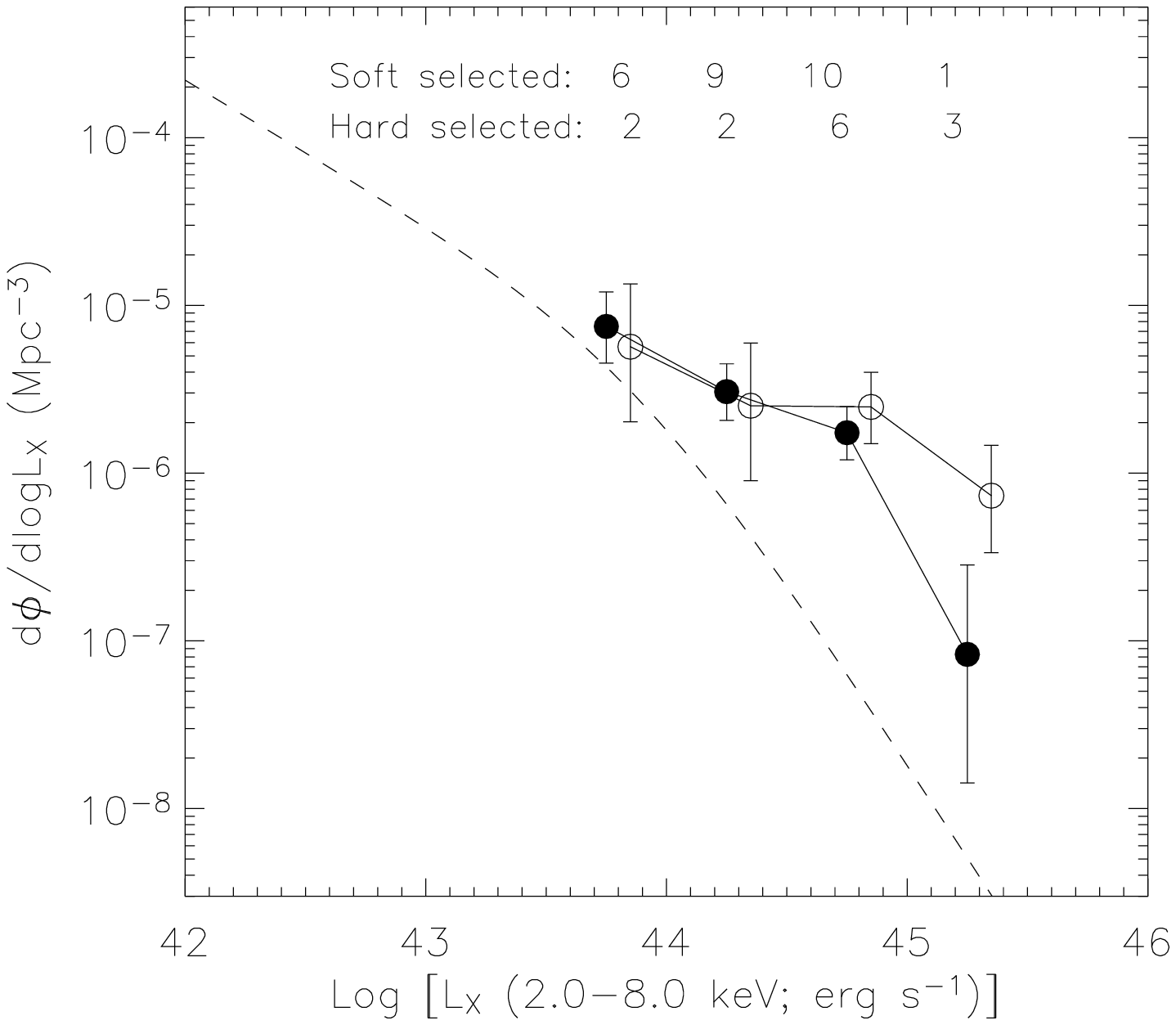}
\caption{XLF in the redshift range $3<z<4$ for AGN selected in the
soft band (filled circles) and hard band (open circles).  The numbers
of AGN in each bin are reported at the top.  The similarity of these
XLFs and the improved statistics in the soft band illustrate our
justification for using an energy dependent selection function.  For
reference, we show the analytic model from \citet{ue03} at $z=0$
(dashed line).}
\label{fig:xlf_compare}
\end{figure}

As previously mentioned, we see a significant drop in the LF above
$z\sim3$ in either normalization or characteristic luminosity over two
redshift intervals (Figure~\ref{fig:xlf}).  This behavior is similar
to the decline seen in the soft and broad {\em Chandra} bands that we
reported in \citet{si05b}. With this new sample, we can further
constrain the slope and normalization of the XLF at these high
redshifts though still larger samples are required especially at $z>4$
where only 4 AGN are included in this analysis.  Between $3<z<4$, our
sample of 26 AGN does show a significant decline from the peak
activity at $2<z<3$.  To check the integrity of our method that
utilizes the observed soft X-ray band to measure the XLF above $z=3$,
we have measured the XLF in the redshift interval $3<z<4$ using the
hard band as done at lower redshifts.  These two measures, shown in
Figure~\ref{fig:xlf_compare}, agree within $\sim1\sigma$ for each data
point.

\begin{figure*}
\hspace{1cm}
\includegraphics[angle=90,scale=0.60]{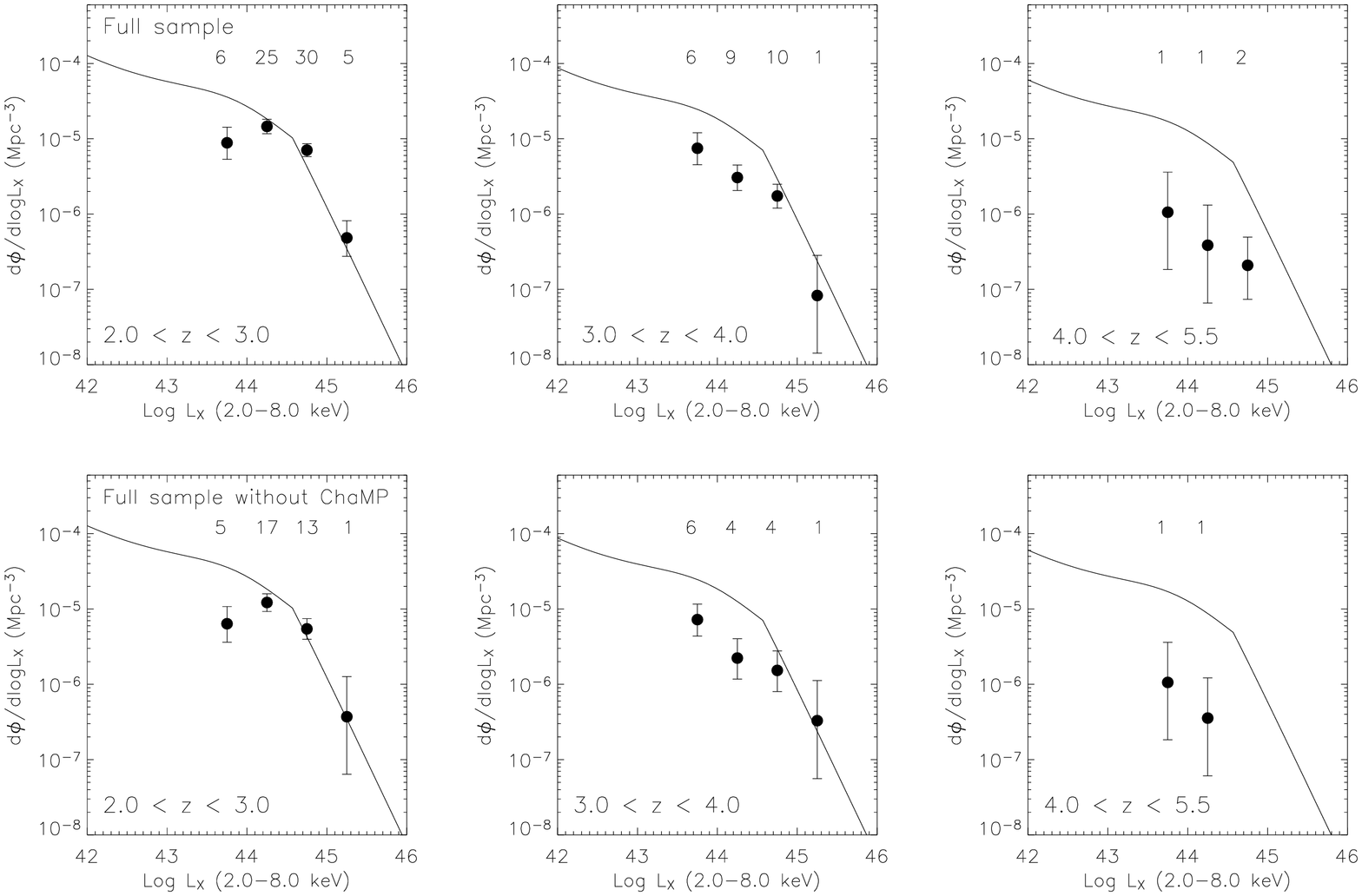}
\caption{A comparison of the high redshift XLF between the full sample
($top$) and that excluding the ChaMP AGN ($bottom$).  The data points
are the same as in Figure~\ref{fig:xlf}.  The solid line is the
best-fit LDDE model of \citet{ue03} that exemplifies our differences
at high redshift.  The number of AGN per luminosity and redshift bin
is shown above their respective data point.}
\label{fig:xlf_nochamp}
\end{figure*}

It is useful to illustrate how the ChaMP sample improves the measure
of the hard XLF.  In Figure~\ref{fig:xlf_nochamp}, we display the XLF
as measured with the $1/V_a$ method in the three highest redshift
intervals.  In the top row, the entire sample is used while the bottom
row includes all AGN with the exception of those from ChaMP.  The
sample size is given for each data point.  For reference, the best-fit
LDDE model of \citet{ue03} (solid line) clearly demonstrates the
difference in our results at $z>3$.  First, it is evident that a
similar decline in the normalization of the XLF is seen with and
without the ChaMP data at $z>3$.  The XLF in the bottom panels is in
agreement with the behavior of the XLF reported by \citet{ba05} using
practically similar samples.  The ChaMP does improve upon the accuracy
of the high redshift XLF by boosting the overall numbers including
those in specific regions of $L_X-z$ plane having limited statistics:
(1) the bright ($log~L_X > 45$) end slope at $2<z<3$ is now further
constrained, (2) measurement errors in the redshift interval $3<z<4$
are effectively reduced, and (3) the numbers of $z>4$ AGN are doubled
with an additional data point at higher luminosities
($log~L_X\sim44.8$) that further constrains the slope.  Results based
solely on the ChaMP sample are presented in both \citet{si05b} and the
PhD thesis of \citet{si04}.

\begin{figure*}
\epsscale{0.8}
\plotone{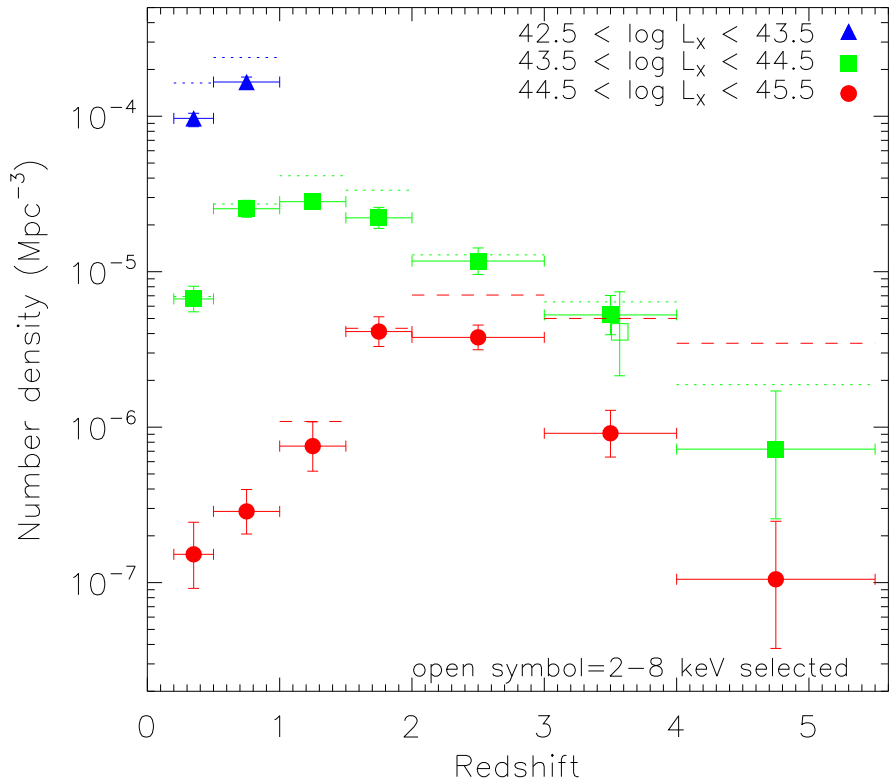}

\caption{Co-moving space density for three luminosity ranges using the
$1/V_a$ method.  The open points show the measured values using the
observed hard band.  The lines (dotted and dashed) illustrate the
maximum uncertainty by assuming that all possible unidentified objects
fall in their respective bins.}

\label{fig:space_density_bin}
\end{figure*}

As frequently represented, the XLF can also be plotted as a function
of redshift for AGN within a fixed luminosity interval.  We show in
Figure~\ref{fig:space_density_bin} the co-moving space density for
three luminosity ranges.  The lowest luminosity range is limited to
$z<1$ due to the lack of X-ray sensitivity at higher redshifts.  For
higher luminosities ($log~L_X>43.5$), our sample provides suitable
numbers of AGN out to $z\sim5$.  We clearly see an increase in the
number of AGN with redshift compared to the local universe followed by
a decline beyond a peak whose redshift is dependent on luminosity.
The most luminous AGN ($L_X>44.5$) peak at a redshift around $z\sim2$,
similar to the optical surveys, while the lower luminosity AGN are
most prevalent at $z\sim1.0$.  We have plotted the measurements using
the observed hard band (open symbols) to show that the decline at
$z>3$ is not due to a significant population of high redshift sources
not detected in the soft band.  These results are in agreement with
similar studies \citep{ue03,ba05,laf05,ha05} with an improvement in
constraints at $z>3$ for the hard XLF.

We note that X-ray sources without redshifts contribute to the
uncertainty in both our XLF and space density.  While those which were
never observed spectroscopically might be expected to be similar in
properties (e.g., in distributions of object type and redshift), the
(significantly smaller) number of objects that we have observed
spectroscopically without achieving a redshift may differ, which could
bias our simple fractional incompleteness correction scheme.  To
illustrate their possible influence, we measure maximum values of the
co-moving space density (Figure~\ref{fig:space_density_bin}) by
placing all possible unidentified objects into each redshift bin
\cite{ba05,si05b}.  We see that the highest redshift bins ($z>3$) are
most susceptible to additional uncertainty if the unidentified objects
have a redshift distribution dissimilar to our AGN sample.

\subsection{Analytic model fitting}

Before we present a global fit over all redshifts and luminosities
spanned by our sample, we verify that our method as described in
Section~\ref{ml_proc} is robust. This is done by fitting our data with
well known models over a similar range of redshift and then comparing
the resulting best-fit parameters to published values.  First, we ran
our ML routine over a redshift interval of $0.2<z<3.0$ using a PLE
model, similar to \citet{ba05} and a LDDE model with fixed parameters
($e2$, $L_a$, and $z_c$) having the same value as given in
\citet{ue03}.  Specifically for the LDDE model, we are interested in
comparing the values of our free parameters descriptive of the shape
($\gamma1$, $\gamma2$), evolution ($e1$) below the cutoff redshift
($z_c$), and strength ($\alpha$) of the dependence of $z_c$ on
luminosity.

In Figure~\ref{xlf_comp}, we plot the best-fit results from our ML
routine ($top$=PLE Model A; $bottom$=LDDE Model B; see
Table~\ref{ml_table} for actual best-fit parameters) and compare to
published results \citep{ba05,laf05,ue03} at $z<3$.  Three redshift
intervals are shown in each case.  Our PLE model ($top$ panel) fit, in
general, agrees with that of \citet{ba05} at $z<1.5$.  Our fit at the
faint end for $z>1$ does not smoothly transverse the binned data
points thus does not adequately represent the slope below the break.
Overall, there is a good agreement between our LDDE model (B; $bottom$
panel) and that of \citet{ue03} and \citet{laf05} with some minor
differences at the faint end that are worth noting.  Below the knee of
the luminosity function, our best-fit model falls below that of these
two aforementioned measures.  As previously noted, this may be a
result of our optical magnitude limits that remove a significant
fraction of the hard and soft X-ray sources above our chosen X-ray
limits.  This may explain (1) our lower normalization ($A_{\circ}$),
though sensitive to the values of the other parameters, as compared to
both the Ueda and La Franca models, and (2) the differences between
the faint-end slope at $z>1$.  The evolution factor ($e1$) and
powerlaw index for the luminosity dependent cutoff redshift ($\alpha$)
are within the 1$\sigma$ errors reported by \citet{ue03}.  Given these
small discrepancies and the limitations of the data, we conclude that
all three models are equally valid at $z<3$.

We aim to extend the fit of the LDDE model to higher redshift ($z>3$).
This has been attempted by \citet{laf05} though with a smaller sample
at these redshifts that may also be biased by cosmic variance since
more than half (5 of 9) of the high redshift AGN are found in the
limited area coverage of the CDF-S.  Since the size our sample is also
restrictive at these redshifts (31), we need to reduce the number of
free parameters.  To do so, we fix those that are highly constrained
at lower redshift: $\gamma1$, $\gamma2$, $L_{\circ}$, and $L_a$.  The
cutoff redshift ($z_c$), evolution parameter above the cutoff redshift
(e2) and $\alpha$ are free to vary.  The most likely fit over the full
redshift range $0.2<z<5.5$ is shown in Figure~\ref{fig:xlf} (dotted
line) with the values of each parameters given in Table~\ref{ml_table}
(Model C).  First, there is agreement with the binned values from the
$1/V_a$ method with some exceptions elaborated further below.  The
main improvement, as compared to published hard XLFs, is that the fit
requires a much stronger evolution ($e2=-3.27_{-0.34}^{+0.31}$) above
the cutoff redshift than previously found by \citet[][$e2=-1.5$]{ue03}
and \citet[][$e2=-1.15$]{laf05}.  Our value is more constrained as
evident by the $1\sigma$ errors that are $\sim2\times$ smaller than
those reported by \citet{laf05}.
  
There is a slight discrepancy between the LDDE fit and the binned data
from our $1/V_a$ method worth highlighting.  The strong evolution
($e2$) above the cutoff redshift from our fit causes the faint end
slope ($log~L_X<44.5$) to flatten a bit more rapidly with increasing
redshift than shown by the binned data.  This is evident in the
$2<z<3$ and $3<z<4$ redshift bins in Figure~\ref{fig:xlf} with the
lowest luminosity points falling above the LDDE fit (dotted line);
although, the statistics are too low to definitely make such claims.
This may just illustrate the difficulties in accurately modelling the
XLF with underlying complexities in addition to the abrupt change from
positive to negative evolution.  It is also worth recognizing that the
faint-end slope of the luminosity function at these high redshifts in
not well constrained since only the CDF-N and CDF-S are capable of
detecting these AGN and their optical identification is quite
challenging.  Some progress has been made using type I AGN selected in
the soft X-ray band \citep{ha05} and optically-selected QSOs found in
Lyman break galaxy surveys \citep{hu04} that clearly show a flatter
faint-end slope compared to lower redshifts.

Finally, we use a modified PLE (mod-PLE) function as described above
and in Table~\ref{ml_table} to allow us greater flexibility.
Accounting for the flattening of the faint-end slope ($\alpha$)
without a dependence on the evolution ($e2$) above the cutoff redshift
may mitigate the problem mentioned above.  The results shown in
Figure~\ref{fig:xlf} (solid line) are also consistent with our binned
$1/V_a$ measurements and show better agreement in the redshift
interval $2<z<3$ than the LDDE fit.  We find the dependence of the
faint-end slope on redshift ($\alpha=-1.04_{-0.12}^{+0.11}$) a bit
stronger than that measured by
\citet[$k_{\gamma1}=-0.623\pm{0.132}$]{ho07}.  At higher redshifts
($z>3$), this model may overestimate the numbers of low luminosity AGN
$log~L_X<44$.  This is evident by comparing the observed and expected
numbers of high redshift AGN in the deep {\em Chandra} Fields (see
Section~\ref{predict}).  Interestingly, the difference between these
two models below the break luminosity becomes larger at $z>4$
with $\sim4\times$ more AGN with $log~L_X\sim43$.  This region of
$L_{X}-z$ parameter space is primarily an extrapolation of the data.
In Section~\ref{predict}, we investigate whether it is possible to
accurately measure the faint-end slope at these high redshifts with
current or future {\em Chandra} observations.

\section{Missed sources, type 2 QSOs and the cosmic X-ray background}
\label{text:cxrb}

The intensity of the Cosmic X-Ray Background, the integrated measure
of the X-ray emission over all redshifts and luminosities, allows us
to assess what fraction of the population we may be missing from our
optically-identified point sources in the 2--8 keV band.  In addition
to constraining an undetected population of X-ray sources, we can also
assess whether our incompleteness corrections to account for
unidentified sources, are reasonable.  The intensity of the CXRB in
the 2--8 keV band has been measured with several hard X-ray
observatories over the last few decades \citep[e.g.,][]{mo03} and has
been shown to have significant scatter (~$\sim20\%$) presumably due to
instrument calibration.  Here, we compare to the best-fit value from
\citet{mo03} of \hbox{$1.79\times10^{-11}$ erg cm$^{-2}$ s$^{-1}$
deg$^{-2}$} in the 2--8 keV band that has been converted from the
2--10 keV band based on an assumed CXRB spectrum with photon index of
1.4.

\begin{figure}
\epsscale{1.1}
\plotone{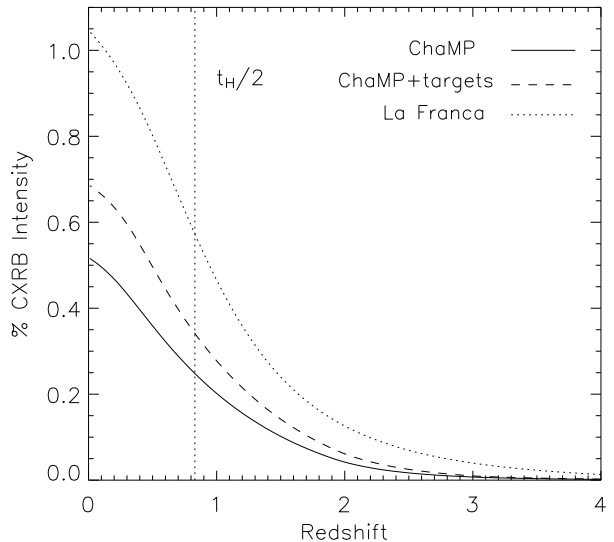}
\caption{Cumulative fraction of the 2--8 keV CXRB as a function of
redshift.  The solid line corresponds to our best-fit LDDE model (C)
and a corrected version (dashed line) that accounts for excluded,
faint ($r^{\prime}, i^{\prime}>24$) targets .  For comparison, we show
(dotted curve) the resolved fraction for the best fit LDDE model of
\citet{laf05}.  The vertical line marks half a Hubble time.}
\label{cxrb}
\end{figure}

In Figure~\ref{cxrb}, we show the integrated contribution of AGN from
our best-fit LDDE model (C; solid line) to the CXRB as a function of
redshift.  As shown, we account for 52\% of the 2--8 keV CXRB at
$z=0$.  We assess the contribution of AGN that have X-ray fluxes above
our limits though fell out of the sample due to their faint optical
magnitudes ($r^{\prime}, i^{\prime}>24$) by applying a correction
based on the assumption that their redshift distribution is equivalent
to identified sources at similar X-ray fluxes.  These sources, as
shown in the top panels of Figures ~\ref{fig:select_hard} and
~\ref{fig:select_soft} by the dot-dashed line, mainly occupy the faint
end of the X-ray flux distribution.  Our measure of the resolved
fraction of the CXRB now reaches $68\%$ (Fig.~\ref{cxrb}; dashed
curve).  These results are not surprising since \citet{wo05} find that
$\sim25$\% of the hard CXRB remains to be resolved.  This is also
consistent with recent resolved fration of $79\pm8\%$ measured in the
2--8 keV band by \citet{hi06}.  Both studies attribute the bulk of
this unresolved emission to be from absorbed AGN (log $N_H\sim23-24$
cm$^{-2}$) with redshift roughly between 0.5 and 1.5.  Therefore, at
most, we may be inaccurately accounting for $\sim10\%$ of the AGN
population due to our rather simple assumptions for the
optically-faint X-ray detections.  Both \citet{ue03} and
\citet{laf05} have measured resolved fractions close to unity.  Other
sources of uncertainty may be due to the fact that we did not correct
our fluxes for X-ray absorption and excluded the targets of each ChaMP
field for which a few are bright AGN \citep{ki07b}.  We further
illustrate in Figure~\ref{cxrb}, that the hard CXRB is mainly
generated by AGN at $z<1$, when the universe was more than half its
present age, as previously described in many works to date
\citep[e.g., ][]{ba01}.

A current limitation of the ChaMP sample is the lack of absorbed
($N_H>10^{22}$ cm$^{-2}$) and luminous ($log~L_X>44$) AGN mainly due
to the bright optical magnitude cut for optical spectroscopic followup
\citep{si05a,gr04}.  The lack of these AGN may not greatly alter our
current luminosity function since they do not seem to outnumber the
unabsorbed (i.e., type 1) AGN at these high luminosities.
\citet{ue03} find the absorbed fraction at these luminosities to be
$\sim$30\% while \citet{laf05} find an even lower fraction of
$\sim20\%$.  An accurate assessment is still required since the
numbers of highly absorbed and luminous AGN (i.e., type 2 QSOs) are
still limited.  The ChaMP project has initiated a deep optical
spectroscopic campaign on Gemini to identify the optically-faint X-ray
sources \citep[see Fig. 13 of][]{si05a} that may provide a fair sample
of these AGN thus removing our current selection bias.  If we do relax
our definition of a type II QSO to having a FWHM < 2000 km s$^{-1}$,
as done by many studies \citep{sz04,ec06}, we have found two such QSOs
at $z>3$ (CXOMP J114036.2+661317, CXOMP J001756.6+163006) as shown in
Figure~\ref{examples}.  The SEXSI survey \citep{ec06,ec05,ha03}, with
an unprecedented sample of 33 luminous AGN lacking broad optical
emission lines, has demonstrated that deep 8-10m class spectroscopy is
key to identify these optically faint AGN.

\section{Comparison with optically-selected AGN}
\label{text:compare_opt}

It is useful to directly compare our XLF to the latest optical
luminosity functions, as also done in recent works
\citep[e.g.,][]{ue03,ba05,ri06}, that cover a similar redshift range.
Currently, the SDSS \citep{ri06,ji06} and COMBO-17 \citep{wo03}, that
solely detect type 1 AGN, cover a redshift range out to $z\sim5$.  We
elect to use cgs units throughout this work and transform the optical
luminosity functions to match our XLF.  We convert the luminosity used
for the SDSS in units of absolute magnitude ($M_i; z=2$) into a
monochromatic luminosity ($l_{2500{\rm \AA}}$) using the
transformation given in Equation 4 of \citet{ri06}.  The COMBO-17
rest-frame magnitudes $M_{1450}$ are converted to $l_{2500{\rm \AA}}$
assuming a powerlaw slope for the spectral energy distribution
($f_{\nu}\propto\nu^{-\alpha}$) with spectral index ($\alpha$) equal
to 0.5.  These monochromatic rest-frame UV luminosities are translated
into a monochromatic X-ray luminosity ($l_{2~{\rm keV}}$) using
Equation 1c of \citet{st06}.  This relation has been shown by many
X-ray-to-optical studies of AGN \citep[e.g.,][]{wi94,gr95,st06} to
have no redshift dependence and is applicable out to $z=5$
\citep{vi05}.  The slope of this relation is further supported, as
shown by \citet{ue03}, by shifting the hard XLF to match the optical
at lower redshifts ($z<2.3$).  We determine the rest-frame 2--8 keV
X-ray luminosity from $l_{{\rm 2 keV}}$ by assuming a powerlaw SED
($L_E \propto E^{-\Gamma}$) with $\Gamma=2.0$ \citep{re00}.

\begin{figure*}
\hspace{3cm}
\epsscale{1.0}
\includegraphics[angle=0,scale=0.75]{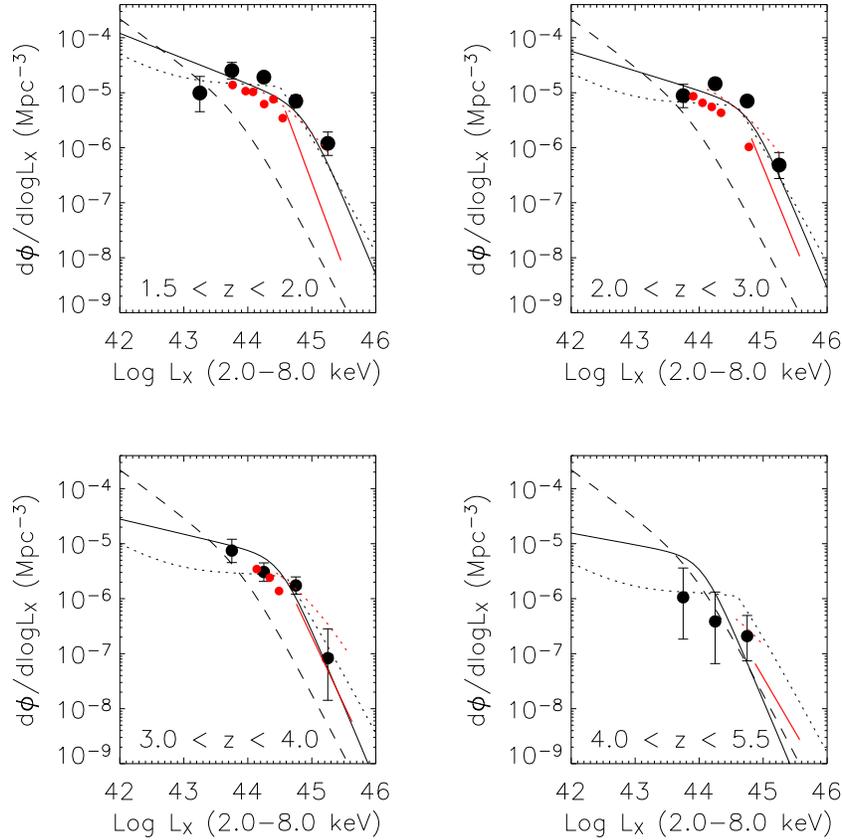}
\caption{Comparison of the X-ray and optical luminosity functions at
$z>1.5$.  The black points and lines are our data as shown in
Figure~\ref{fig:xlf}.  The optical luminosity function is overplotted
for both the SDSS \citep[red line;][]{ri06}, optically-faint SDSS
\citep[small red dots;][]{ji06}, and COMBO-17 \citep[dotted red
line;][]{wo03}.}
\label{fig:xolf}
\end{figure*}

We show in Figure~\ref{fig:xolf} a comparison of our X-ray luminosity
function with the data and analytic model fits from the SDSS and
COMBO-17 in the four highest redshift intervals mainly to highlight
the behavior above and below the QSO peak at $z\sim2.5$.  For the
optically-selected luminosity functions, we limit the luminosity range
to that which is covered by these surveys.  This allows us to
illustrate the complementarity of these surveys and avoid any
interpretation based on extrapolations beyond the dynamic range of
each sample.  In general, the bright end slope of our XLF agrees with
that from the optical surveys out to $z\sim5$ with the exception of
the $1.5<z<2.0$ redshift interval where the SDSS model is much steeper
even compared to both X-rays and to COMBO-17 \citep[see][]{ri06}.
Qualitatively, there does not appear to be any strong evidence for a
change in the slope of the bright end with redshift (see
Figure~\ref{fig:xlf} for a larger redshift baseline), as shown
recently \citep{ri06,ho07}. At lower redshifts ($z<1.5$), the slope at
the bright end is not nearly as steep as measured in \citet{ri06},
though larger samples of luminous X-ray selected AGN are required to
justify such a statement.  As shown, the faint end of the luminosity
function can be probed by optically-selected surveys \citep{ji06,hu04}
of type 1 AGN at $z<3$.  As expected, their space density falls below
those found from X-ray selection due to the lack of obscured AGN.  At
$z>3$, the luminosities of these optically-selected AGN do not fall
well below the knee.  As evident, X-ray surveys can probe
moderate-luminosity AGN at $3<z<4$, fainter than the break luminosity
($log~L_X<44.5$), thus allowing a measure the slope at the faint end.
Some uncertainty remains since the contribution of luminous type 2
QSOs is still uncertain especially at $z>3$.  However, as previously
mentioned, recent constraints on the type 2 QSO population
\citep[$\sim$25--50$\%$][]{gi07} tend to show limited effect upon our
current measure of the XLF even when considering the possible increase
in the obscured fraction with redshift \citep{tr06} since the
luminosity dependence is stronger than the redshift dependence.

\section{Accretion history of SMBHs}
\label{history}

The luminosity function of AGN is one of the key observational
constraints on the mass buildup of SMBHs over cosmological time.  With
our measure of the XLF less biased against obscuration, sensitivity to
lower luminosity AGN and coverage out to high redshift, we can
reexamine, with very simplistic assumptions, our understanding of how
the mass accretion rates, Eddington masses and the cumulative mass
density of SMBHs behaves as a function of redshift.  In addition, we
can ascertain whether the well known similarity between the accretion
rate onto SMBHs and the star formation history of galaxies
\citep{bo98,fr99,me04b}, from the QSO peak ($z\sim2.5$) to the
present, persists at higher redshifts.

We determine the distribution of mass accretion rate as a function of
redshift with the assumption that the accretion rate is directly
proportional to the bolometric luminosity ($L_{BOL}$;
Equation~\ref{eq:maccr}) by a constant mass to energy conversion
factor $\eff_{rad}$.

\begin{equation}
L_{BOL}= \eff_{rad} \Mdot_\mathrm{acc} c^2
\label{eq:maccr}
\end{equation}

\begin{figure}
\epsscale{1.1}
\plotone{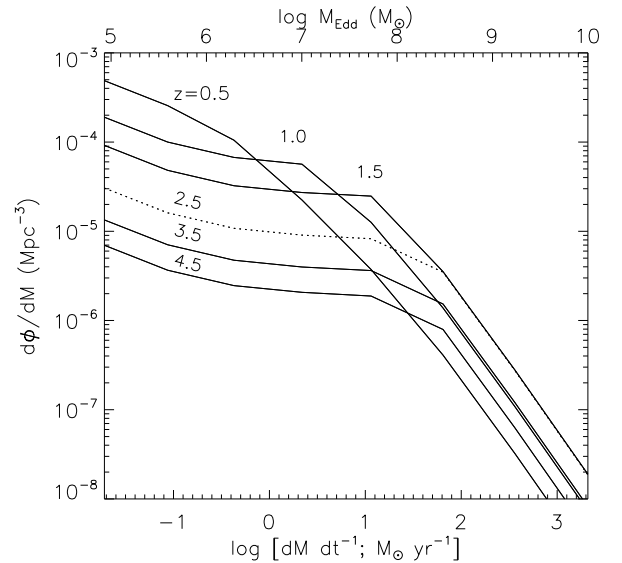}
\caption{Distribution of mass accretion rate onto SMBHs at redshifts
above and below the peak number density ($z=2.5$; dashed line) using
our best-fit LDDE model (C) and a fixed accretion efficiency.  The top
axis shows the Eddington masses assuming an equivalent bolometric
luminosity.}
\label{fig:bh_physics}
\end{figure}

\noindent We derive bolometric luminosities from the rest-frame 2--8
keV X-ray luminosity using the non-linear relation given in Equation
21 of \citet{ma04} that mainly accounts for changes in the overall
spectral energy distribution of AGN as a function of optical
luminosity \citep[e.g.,][]{st06}.  We fix the accretion efficiency
$\eff_{rad}$ to 0.1 \citep[e.g.,][]{ma04}.  In
Figure~\ref{fig:bh_physics}, we plot the accretion rate distribution,
using our best-fit LDDE model (C), at six redshifts that span the full
extent of our XLF.  Using the same bolometric luminosities, we label
the top axis with the corresponding Eddington masses.  We stress that
these accretion rates and masses are highly degenerate.  Here, we aim
to just illustrate the evolutionary behavior of the population in
terms of their physical properties rather than present
well-constrained physical measurements.  In light of these words of
caution, we see that the number density of the entire distribution
systematically rises with cosmic time from $z=4.5$ to $z=2.5$ (dotted
line) that may represent the global ignition of mass accretion onto
SMBHs at early epochs.  At redshifts below this peak, the number
density of AGN with high accretion rates (log$~\Mdot\gtrsim1$) or
masses (M$_{Edd}\gtrsim10^{8}$ \Msun) essentially halts to $z\sim1.5$
and then declines rapidly.  In contrast, the number of AGN with low
accretion rates ($log~\Mdot\lesssim1$) or masses
(M$_{Edd}\lesssim10^{7}$ \Msun), continues to climb.  Either, we are
witnessing the emergence of young SMBHs (i.e., low mass, high
accretion rate) or the remnants of a once highly luminous SMBH
population (i.e., high mass, low accretion rate).  Much recent
evidence points to a luminosity-dependent lifetime
\citep[e.g.,][]{ho05,ad05} that must also be folded in to accurately
extract physical quantities from a luminosity function.  Evidently, a
more careful deconvolution is warranted.

\begin{figure}
\epsscale{1.8}
\plottwo{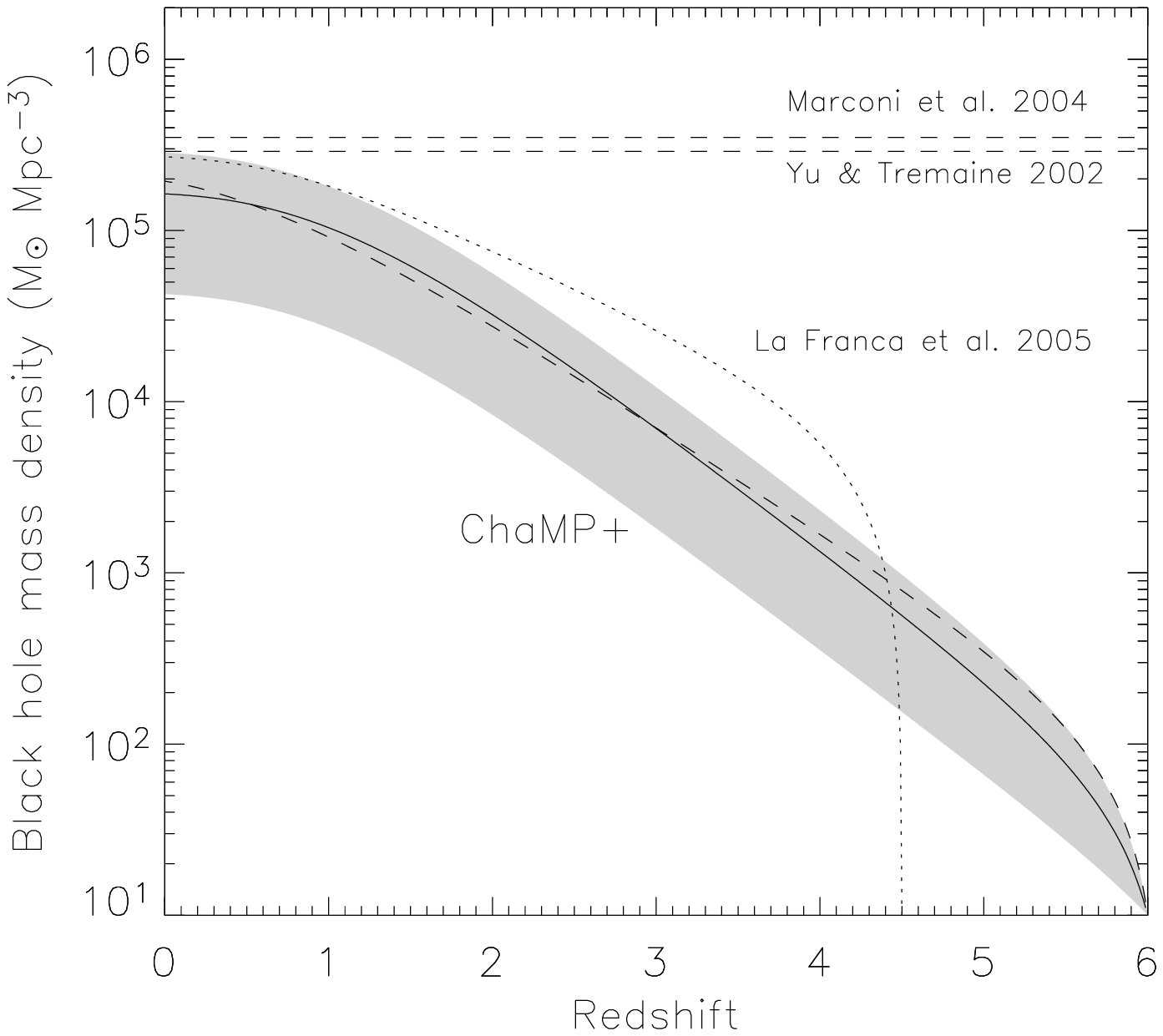}{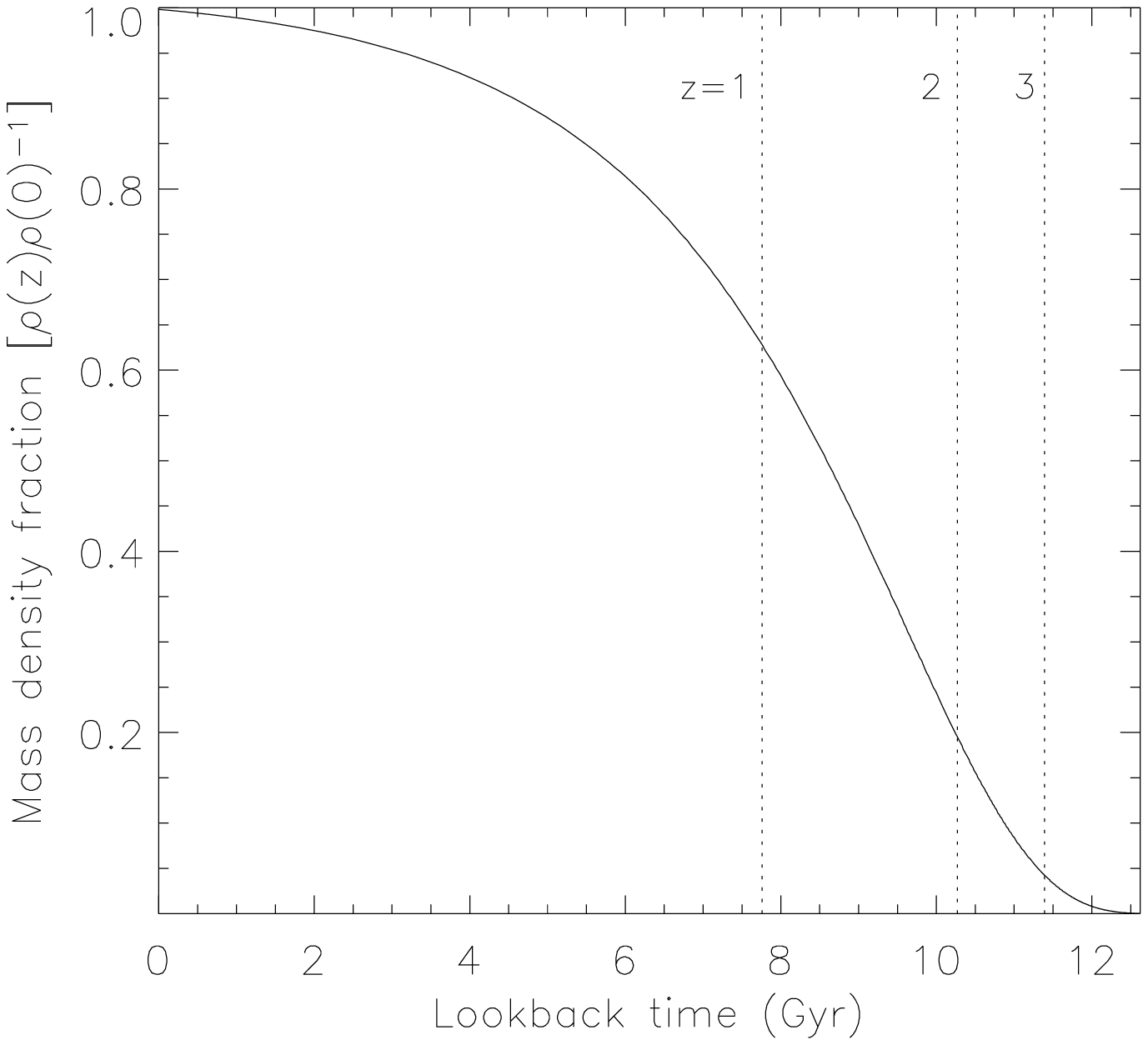}

\caption{Integrated mass density of SMBHs.  $Left$ The dark grey
region shows our cumulative values using our best fit mod-PLE model
(D), a luminosity-dependent bolometric correction, and a range of
accretion efficiencies of $0.06 <\epsilon_r <0.3$ with a solid line
marking the $\epsilon_r=0.1$ case.  We plot with a dashed line the
results using a constant bolometric correction and $\epsilon_r=0.1$.
For comparison, we have highlighted the $z=0$ mass densities
(horizontal, dashed lines) measured using the optical QSO luminosity
function \citep{yu02} and hard XLF \citep{ma04} of \citet{ue03}.  The
dotted line marks the mass density as a function of redshift using the
hard XLF of \citet{laf05}.  $Right$ We show the cumulative mass density,
identical to the case on the left marked by a solid line, as a
function of lookback time normalized by our local value.}
\label{bhgrowth}
\end{figure}

As done in many recent studies, we can measure the cumulative mass
density of SMBHs as a function of redshift from the integrated
light from AGN \citep{so82} using the same assumption for the
radiative efficiency given above.  In particular, the local value
($z=0$) allows a direct comparison with complementary techniques
(e.g., local galaxy luminosity function and a relation between black
hole mass and its host bulge; integrated light from optically-selected
QSOs).  With the dropoff in the number of AGN at $z>3$, we do not
expect our measure of the local value of the mass density to differ
from recent estimates but aim to illustrate that further constraints
on the mean growth of AGN at high redshifts are now possible.  In
Figure~\ref{bhgrowth}, we plot the results using our best fit
'mod-PLE' model (D) over an extended luminosity
($41<log~L_{2-8~{\rm keV}}<47$) and redshift range ($0<z<6$).  We do not
account, as done in Section~\ref{text:cxrb}, for the optically-faint
X-ray sources not included in our sample; their contribution is not
trivial to assess since their redshift distribution is most likely
dissimilar to optically-brighter AGN at similar X-ray fluxes as
demonstrated by \citet{ma05}.  No correction for intrinsic absorption
is applied.  We have measured the mass density of black holes for two
extreme values of the radiative accretion efficiency that is less
constrained than the bolometric corrections involved.  The grey area
illustrates the range of mass densities with values of the accretion
efficiency ranging from a Schwarzschild ($\epsilon_r=0.06$) to a Kerr
($\epsilon_r=0.3$) black hole.  The solid line gives the values for
the widely accepted value $\epsilon=0.1$ with a mass density at $z=0$
of $1.64\times10^5$ \Msun~Mpc$^{-3}$.  In addition, we show the value
for a constant bolometric correction \citep[40;][]{elvis94} with
$\epsilon=0.1$.  Our estimate of the local mass density in SMBHs is
lower than other measures ($left$).  Our value is 48\% of the the
local value, determined from the galaxy luminosity function and
velocity disperson, of $3.4_{-0.5}^{+0.6}\times10^5$ \Msun~Mpc$^{-3}$
\citep{ma04} and is in closer agreement to the values obtained from
previous hard X-ray-selected AGN \citep{ba05,laf05,ue03}.  In
particular, \citet{ba05} find a similar local black hole mass density
($2.1\times10^5$ \Msun~Mpc$^{-3}$) when using a bolometric correction
factor (35) close to that employed here.  \citet{ba05} point out that
the mass density measured by \citet{yu02} is possibly too high due to
an extrapolation of the optical luminosity function at the faint end
that has now been shown to be shallower \citep{cr04}.  The fact that
we are not accounting for optically-faint X-ray sources, resolving
$\sim$70\% of the hard CXRB, and do not account for Compton-thick AGN
could contribute to the discrepancy with mass density estimates using
local galaxies.  As is evident in Figure~\ref{bhgrowth} $right$, most
(81\%) of the growth of SMBHs occured by the time the universe was
half ($z\sim0.7$) its present age with minimal contribution at $z>3$.

\subsection{Co-evolution of SMBHs and star-forming galaxies}
\label{coevolution}

The order of magnitude decline in the co-moving mean emissivity of
AGN and star formation history of galaxies from $z\sim1.5$ to the
present has initiated many early postulates for a co-evolution scheme
of SMBHs and the galaxies in which they reside.  \citet{bo98}
demonstrated that the optical QSO luminosity function exhibited a
similar behavior to the star formation rate of field galaxies.
\citet{fr99}, with an X-ray selected sample of AGN, further
investigated this correlation by demonstrating that the more luminous
QSO evolve closely with the most massive galaxies (i.e., E/S0) while
the lower-luminosity AGN track the star-forming, field galaxy
population.  At the time, these studies were mainly limited to $z<3$
due to the availability of measured SFRs of galaxies.

Armed with our XLF of AGN out to $z\sim5$ and the latest SFRs of high
redshift galaxies, we can revisit this correlation.  In the last few
years, new samples of high redshift galaxies has enabled the evolution
of the SFR to be extended out to $z\sim6$
\citep[e.g.,][]{bu04,gi04,bo07}.  Most studies have selected these high
redshift galaxies by Lyman break techniques \citep[e.g.,][]{st99} or
narrow emission line searches \citep[e.g.,][]{rh03}.  With the advent
of wide area, multi-slit spectrographs on 8--10m class telescopes,
flux-limited surveys are able to probe a sizeable volume and identify
significant numbers of high redshift galaxies without preset color
selection criteria or emission line strengths \citep{lef05}.

\begin{figure}
\epsscale{1.1}
\plotone{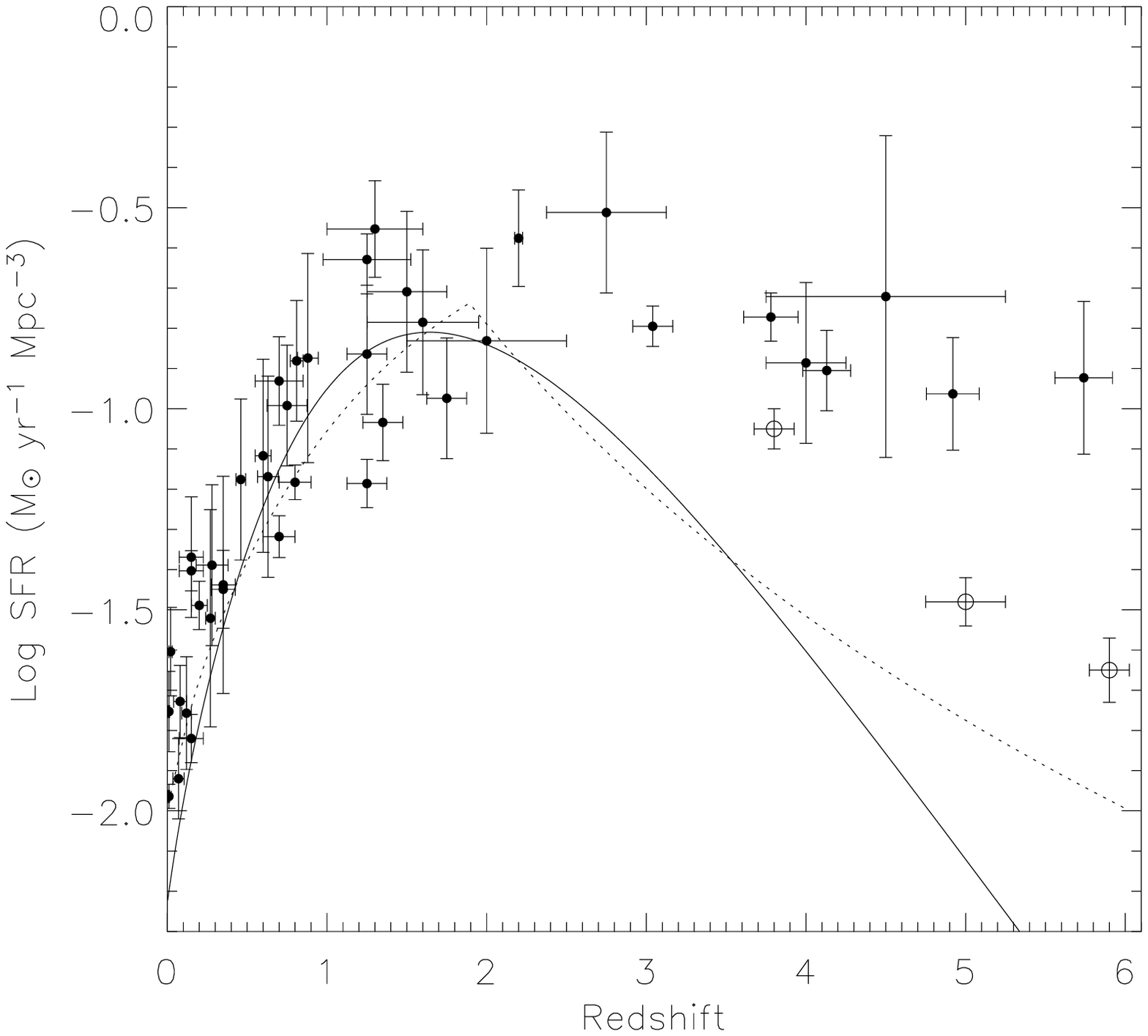}
\caption{Comparison of the star formation history of galaxies (data
points) to the mass accretion rate ($\Msun$ yr$^{-1}$ Mpc$^{-3}$) of
SMBHs using the Mod-PLE (solid line) and LDDE (dotted line) models.
SFRs from the compilation of \citet{ho04} are marked by the small
filled dots while the open circles are recent measurements at
$z\gtrsim4$ \citep{bo07}.  The mass accretion rates have been scaled
up by a factor of 5000.  The divergence at $z>4$ of the two curves
representing the accretion history of SMBHs highlights the poor
constraints on the X-ray luminosity function at early epochs.}
\label{coevol}
\end{figure}

We have measured the mass accretion rate per co-moving volume (units
of $\Msun$ yr$^{-1}$ Mpc$^{-3}$) as a function of redshift using our
analytic models (C, D) for all AGN with $L_{2-8~{\rm keV}}>10^{42}$
erg s$^{-1}$.  Bolometric corrections and conversion factors to
accretion rates are used as described in the previous section.  In
Figure~\ref{coevol}, we plot the results compared to the
dust-corrected SFRs (black data points) as compiled by \citet{ho04}.
We add the latest measurements at $z\gtrsim4$ from \citet{bo07} as
shown by the open circles.  The mass accretion rates have been scaled
up by a factor of 5000.  Up to $z\sim2$, there is an overall
similarity as previously elaborated in many studies to date.  The
discrepancy noted by \citet{ha04} at $z<2$ appears to be removed with
the inclusion of obscured AGN from X-ray selected surveys.  The peak
in the black hole accretion rate density now occurs at $z\sim1.5$
rather than $z\sim2.5$ as measured with optically-selected QSO
samples.  Also noted in many investigations, there is a divergence at
$z>2$ between the two growth rates, with a faster decline of the AGN
population even when comparing to the recent SFRs \citep{bo07} that
show a significant decline between $4<z<6$.  Here we find a decrease
in the accretion rate onto SMBHs by a factor of 3--4 compared to the
SFRs at $z\sim4.5$.  This may represent a delay between the formation
of stars and fully matured SMBHs in the early universe possibly due to
either the cooling times required for gas to be available for
accretion or an insufficient merger history \citep[See ][for further
discussion]{ha04}.

\section{Detection of high redshift AGN in {\em Chandra} surveys} 
\label{predict}

Even though we have extended the 2--8 keV luminosity function beyond
$z\sim3$, there remains much uncertainty about the behavior of the
luminosity function below the break in the $3<z<4$ redshift range and
at all luminosities at higher redshifts.  Since {\em Chandra} has now
completed or will undertake some new surveys that cover a wider area
and at moderate depths, we can make some predictions of their return
of high redshift AGN.  We have computed the numbers of AGN in two
redshift intervals and two luminosity ranges that basically span below
($42<log~L_X<44$) and above ($44<log~L_X<46$) the break luminosity.
We give these values for various surveys using our best-fit LDDE and
mod-PLE (in parenthesis) models in Table~\ref{table:predict}.  The
estimates between the two models are similar at high luminosities but
differ at the low luminosity end, which becomes even more pronounced at the
highest redshifts (see Figure~\ref{fig:xlf}).

We see that the ChaMP will provide a total of $\sim46$ AGN at $z>3$
based on an area coverage of 2 square degrees that is similar to the
numbers expected from the latest wider area surveys (i.e., EGS, {\em
Chandra}/COSMOS).  As mentioned earlier, these surveys are
complementary to ChaMP since they reach greater depths over a smaller
area to detect AGN at lower luminosities as exemplified here.  We note
that the mod-PLE model is most likely overpredicting the numbers of
lower luminosity ($log~L_X<44$) AGN based on the observed number in
the CDF-N and CDF-S.  It total, we expect a final sample of over
$\sim$195 AGN at $z>3$ with 63 above a redshift of 4. Better
constraints on the faint end of the luminosity function in the $3<z<4$
redshift interval can be achieved by either the E-CDF-S, EGS, or
cCOSMOS with a sample of $\sim15-25$ sources per survey though an
accurate measure of the slope will require the combined sample that is
especially true at higher redshifts.  An accurate assessment of the
luminosity function at $z>4$ will definitely require a merger of all
available catalogs.

\section{Summary}

We have presented an extension of the hard (2--8 keV) X-ray luminosity
function of AGN up to $z\sim5$.  The ChaMP effectively covers a wide
area (1.8 deg$^{2}$) at sufficient depths
\hbox{($f_{0.5-2.0~{\rm keV}}\sim10^{-15}$ erg cm$^{-2}$ s$^{-1}$)} to
significantly improve the statistics of luminous ($L_{\rm X}>10^{44}$ erg
s$^{-1}$) AGN at high redshift.  In total, we have amassed a sample of
682 AGN with 31 at $z>3$.  The addition of lower luminosity AGN from
the {\em Chandra} Deep Fields is instrumental to characterize the faint
end slope at $z\gtrsim1$.  We have corrected for incompleteness
(i.e., fraction of sources without a redshift) as both a function of
X-ray flux and optical magnitude, as dictated by the limitations of
optical spectroscopic followup.  Significant optical
followup is still required to accurately account for the obscured
population, especially at high redshifts.

We have measured the hard XLF with both a binned ($1/V_a$) and an
unbinned method that fits an analytic model to our data using a
maximum likelihood technique.  We found that the luminosity function
is similar to that found in previous studies \citep{ue03,laf05,ba05}
up to $z=3$, with an evolution dependent upon luminosity.  At higher
redshifts, there is a significant decline in the numbers of AGN with
an evolution rate similar to that found with optically-selected QSO
samples.  We further show that the strong evolution above the cutoff
redshift may cause the LDDE model to underpredict the number of low
luminosity AGN at $z>2$.  A PLE model with a faint-end slope dependent
on redshift agrees better with the binned ($1/V_a$) data at $z\sim2.5$
though it may overpredict the number of faint AGN at higher redshifts.
We highlight the need to improve the statistics of both high redshift
AGN at $z>4$ and lower luminosity (i.e., below the break in the
luminosity function) AGN at $z>3$. Such improvements are feasible from
our predictions of the numbers AGN that will be found in the latest
{\em Chandra} surveys (E-CDF-S, EGS and COSMOS)

Our new luminosity function accounts for $\sim52\%$ of the 2--8 keV
Cosmic X-ray Background.  The integrated emission from these AGN give
a $z=0$ mass density of SMBH of $1.64\times10^5$ \Msun~Mpc$^{-3}$,
lower than other published values using X-ray selected AGN samples and
the local value measured using a galaxy luminosity function and a
bulge-velocity relation, possibly due to unaccounted optically-faint
AGN ($r^{\prime},i^{\prime}>24$), and Compton-thick AGN.  Further, a
comparison of the mean mass accretion rate of SMBHs to the star
formation history of galaxies out to $z\sim5$ shows the familiar
co-evolution scheme up to $z\sim2$ and a divergence at higher
redshifts with perhaps star formation preceeding the formation of
SMBHs.

\acknowledgements

The authors wish to thank the NOAO and SAO telescope allocation
committees for their support of this work.  We also recognize the
contributions from the following people for either their insightful
discussions, general comments on the work, or providing electronic
tables such as the XMM-$Newton$ data: Dave Alexander, Hermann Brunner,
Jacobo Carrero, Fabio La Franca, Vincenzo Mainieri, Israel Matute,
Andrea Merloni, Takamitsu Miyaji, Francesco Shankar, and Gyula Szokoly.

Facilities: \facility{CXO},
\facility{Mayall},
\facility{Blanco},
\facility{WIYN},
\facility{Magellan:Baade},
\facility{Magellan:Clay},
\facility{MMT},
\facility{FLWO:1.5m}

\clearpage

\begin{deluxetable*}{llllllll}
\tabletypesize{\scriptsize}
\tablecaption{AGN selected from various X-ray surveys\label{table_surveys}}
\tablewidth{0pt}
\tablehead{
\colhead{Name}&\colhead{Telescope}&\multicolumn{3}{c}{Hard band selection ($z<3$)}&\multicolumn{3}{c}{Soft band selection ($z>3$)}\\
&&\colhead{Depth}&$r^{\prime}$&\colhead{Number}&\colhead{Depth}&$i^{\prime}$&\colhead{Number}\\
&&\colhead{(erg cm$^{-2}$ s$^{-1}$)}&(mag)&\colhead{of AGN}&\colhead{(cgs)}&(mag)&\colhead{of AGN}
}
\startdata
ChaMP&{\em Chandra}&$2.7\times10^{-15}$&22.0&273&$1.0\times10^{-15}$&22.0&13\\
CLASXS&{\em Chandra}&$2.3\times10^{-15}$&24.0&103&$5.6\times10^{-16}$&24.0&3\\
CDF-N&{\em Chandra}&$6.3\times10^{-16}$&24.0&104&$1.0\times10^{-16}$&24.0&8\\
CDF-S&{\em Chandra}&$6.3\times10^{-16}$&24.0&95&$1.0\times10^{-16}$&24.0&2\\
Lockman Hole&XMM-$Newton$&------&------&------&$1.0\times10^{-15}$&24.0&5\\
AMSSn&$ASCA$&$2.52\times10^{-13}$&22.0&76&------&------&------\\
\enddata
\end{deluxetable*}

\begin{deluxetable}{llll}
\tabletypesize{\scriptsize}
\tablecaption{ChaMP fields\label{champ_fields}}
\tablewidth{0pt}
\tablehead{\colhead{Obs. ID}&\colhead{ACIS} & \colhead{PI Target}& \colhead{Exposure}\\
&&&\colhead{time (ksec)}
}
\startdata
520 &    I  &     MS0015.9+1609&61.0\\	
913 &    I  &     CLJ0152.7-1357&34.6\\
796 &    I  &     SBS0335-052&47.0\\
624 &    S  &     LP944-20&40.9\\
902 &    I  &     MS0451.6-0305&41.5\\
914 &    I  &     CLJ0542.8-4100&48.7\\
377 &    S  &     B2 0738+313&26.9\\
2130&    S  &     3C207&30.0		\\
419 &    S  &     RXJ0911.4+0551&30.0  \\
839&	S&	3C220.1&20.0	\\
512&     S &      EMSS1054.5-0321&75.6	\\
363&	S&	PG1115+080&30.0\\
536&     I&       MS1137.5+6625&114.6\\
874&     I&       1156+295&75.0	\\
809&     S&       Mrk237X&50.0\\
541&     I&       V1416+4446&29.8\\
800&	S&	CB 58	&50.0	\\		
546&     I &      MS1621.5+2640&30.0\\
830&     S &      Jet of 3C390.3&23.6\\
551&     I &      MS2053.7-0449&42.3\\
928&     S &      MS2137.3-2353&29.1\\
431&     S &      Einstein Cross&21.9\\
918&     I &      CLJ2302.8+0844&106.1\\
861&     S &      Q2345+007&65.0\\
\enddata
\end{deluxetable}

\begin{deluxetable}{lllllll}
\tabletypesize{\scriptsize}
\tablecaption{ChaMP statistics\label{champ_stats}}
\tablewidth{0pt}
\tablehead{\colhead{X-ray}&\colhead{X-ray flux}&\colhead{Optical}&\multicolumn{3}{c}{Numbers}&\colhead{Id fraction}\\
\colhead{band}&\colhead{limit\tablenotemark{a}}&\colhead{limit}&\colhead{All}&\colhead{Spectra}&\colhead{Ids\tablenotemark{b}}&\colhead{All~~observed}}
\startdata
Hard&$2.7\times10^{-15}$&none&793&456&328&41\%~~~~72\%\\ 
&&$r^{\prime}<24.0$&598&444&325&54\%~~~~73\%\\
&&$r^{\prime}<22.0$&390&332& 289&74\%~~~~87\%\\
Soft&$4.0\times10^{-16}$&none&1125&624&435&39\%~~~~70\%\\
&&$i^{\prime}<24.0$&834&605&430&52\%~~~~71\%\\
&&$i^{\prime}<22.0$&512&429&373&73\%~~~~87\%\\
\enddata
\tablenotetext{a}{units of erg cm$^{-2}$ s$^{-1}$}
\tablenotetext{b}{Quality flag equal to two or three.}
\end{deluxetable}

\clearpage

\begin{landscape}
\begin{deluxetable}{llllllllllll}

\tabletypesize{\scriptsize}
\tablecaption{Hard XLF Best-Fit Model Parameters\label{ml_table}}
\tablewidth{0pt}
\tablehead{
\colhead{Model} &
\colhead{Type} &
\colhead{$log~A_{\circ}$} &
\colhead{$\gamma$1} &
\colhead{$\gamma$2} &
\colhead{$log~L_{\circ}$} &
\colhead{e1} &
\colhead{e2} &
\colhead{$z_c$} &
\colhead{$log~L_a$} &
\colhead{$\alpha$}&
\colhead{Redshift range}
}
\startdata
A&PLE&--5.332$\pm{0.015}$&2.80$\pm{0.20}$&0.77$\pm{0.05}$&44.92$\pm{0.10}$&1.69$\pm{0.12}$&--0.87$\pm{0.6}$&1.9\tablenotemark{a}&------&------&$0.2<z<3.0$\\
B&LDDE&--6.077$\pm{0.015}$&2.15$_{-0.12}^{+0.42}$&1.10$\pm{0.13}$&44.33$\pm{0.10}$&4.00$\pm{0.28}$&--1.5\tablenotemark{a}& 1.9\tablenotemark{a}&44.6\tablenotemark{a}&0.317$_{-0.027}^{+0.020}$&$0.2<z<3.0$\\
C&LDDE&--6.163$\pm{0.015}$& 2.15\tablenotemark{b}&1.10\tablenotemark{b}&44.33\tablenotemark{b,c}&4.22$_{-0.27}^{+0.20}$&--3.27$_{-0.34}^{+0.31}$&1.89$_{-0.06}^{+0.14}$&44.6\tablenotemark{a}& 0.333$\pm{0.013}$&$0.2<z<5.5$\\
D&Mod-PLE&--5.238$\pm{0.015}$&2.76$_{-0.19}^{+0.21}$&0.42$\pm{0.05}$&44.88$_{-0.29}^{+0.11}$&--0.60$_{-0.14}^{+0.15}$&--8.18$\pm{0.55}$&2.0(Fixed)&------&--1.04$_{-0.12}^{+0.11}$&$0.2<z<5.5$\\
\enddata
\tablenotetext{a}{Fixed parameters to those best fit values from \citet{ue03}}
\tablenotetext{b}{Fixed to the best fit value in Model B.}
\tablenotetext{c}{For this case, $L_*=L_0$}
\end{deluxetable}

\begin{deluxetable}{lllllr}
\tabletypesize{\scriptsize}
\tablecaption{Predicted number of high redshift AGN in complete surveys\label{table:predict}}
\tablewidth{0pt}
\tablehead{
\colhead{Survey}&\multicolumn{2}{c}{$3<z<4$}&\multicolumn{2}{c}{$4<z<6$}&Total\\
&\colhead{$log~L_X<44$}&\colhead{$log~L_X>44$}&\colhead{$log~L_X<44$}&\colhead{$log~L_X>44$}\\
}
\startdata
ChaMP&3 (6)       &31 (41) &0 (0)  &12 (6)  & 46 (53)\\
CDF-N&4 (16)      &3 (5)   &2 (9)  &2 (2)  & 11 (32)\\
CDF-S&4 (12)      &3 (4)   &2 (6)  &2 (2)  & 11 (24)\\
E-CDF-S& 11 (35)  &8 (13)  &4 (18) &6 (5)  & 29 (71)\\
EGS& 12 (38)     &16 (25)  &4 (13) &11 (8) & 43 (84)\\
cCOSMOS& 15 (45) &22 (35)  &4 (14) &14 (12) & 55 (106)\\
\hline
Totals&49 (152)&73 (114)&16 (60)&47 (35)& 195 (370)\\
\enddata
\end{deluxetable}
\clearpage

\end{landscape}

\end{document}